\documentclass[aps,prb,reprint]{revtex4-1}
\usepackage{epsfig}
\usepackage{graphicx}
\usepackage{amsmath}
\usepackage{amsfonts}
\usepackage{amssymb}
\usepackage{ulem}
\usepackage{cancel}
\usepackage{graphicx}
 \usepackage{wrapfig}
\usepackage[cp1251]{inputenc}
\usepackage{epsfig}
\usepackage{graphicx}
\usepackage{amsmath}
\usepackage{amsfonts}
\usepackage{amssymb}
\usepackage{graphicx}
 \usepackage{wrapfig}
\usepackage{textcomp}
\usepackage{pgf,pgfarrows,pgfnodes,pgfautomata,pgfheaps,pgfshade}
\usepackage{amsmath,amssymb}
\usepackage[colorlinks=true, linkcolor=blue]{hyperref}
\newcommand{{\sign}}{\rm sign}

\newcommand{\im}{\rm Im}
\newcommand{\diag}{\rm diag}

\begin{document}
\title{
Current-phase relation and $h/e$-periodic critical current of a chiral Josephson contact between 1D Majorana modes}
\author{Dmitriy S. Shapiro$^{1,2,3}$}
\email{shapiro@cplire.ru}
\author{Alexander Shnirman$^{4}$} \author{Alexander D. Mirlin$^{1,4,5}$}
\affiliation{$^1$Institut f\"ur Nanotechnologie, Karlsruhe Institute of Technology, 76021 Karlsruhe, Germany}
\affiliation{$^2$V. A. Kotel'nikov Institute of Radio Engineering and Electronics, Russian Academy of Sciences, Moscow 125009, Russia}
\affiliation{$^3$Dukhov Research Institute of Automatics (VNIIA),  127055, Moscow, Russia}
\affiliation{$^4$Institut f\"ur Theorie der Kondensierten Materie,
Karlsruhe Institute of Technology, 76128 Karlsruhe, Germany}
\affiliation{$^5$Petersburg Nuclear Physics Institute, 188300, St.Petersburg, Russia}

\begin{abstract}
We explore a long Josephson contact transporting Cooper pairs between 1D charge-neutral chiral Majorana modes in the leads via charged Dirac chiral modes in the normal region. We investigate the regimes of (i) transparent contacts and (ii) tunnel junctions implemented in 3D topological insulator/superconductor/magnet hybrid structures. The setup acts as a SQUID controlled by the magnetic flux enclosed by the chiral loop of the normal region. This chirality leads to the fractional $h/e$-periodic pattern of critical current. The current-phase relation can have sawtooth-like shape with spikes at unusual even phases of $2\pi n$.

\end{abstract}

\maketitle
\section{Introduction}

Intensive studies of Majorana fermion physics in solid state materials were motivated by  possible applications of these  states
in topological quantum computation  \cite{Alicea, NayakReview}. In condensed matter, initially,  Majorana fermions were predicted to exist  in spinless  $p$-wave superconductors (SC) and  fractional QHE \cite{Kitaev, ReadGreen, MooreRead}. Later, the realizations of Majorana bound states (MBS) \cite{FuKane,OregRefaelOppen,NilssonAkhmerov} and 1D charge-neutral modes \cite{FuKane3DTI} were proposed in hybrid structures where  helical  electronic system has  a  proximity effect with external $s$-wave pairing potential and Zeeman exchange field. The helical systems with spin and momentum locking  are  realized in surfaces and  edges of topological insulators (TI), quantum spin-Hall films \cite{HasanKane} and  semiconductors with strong spin-orbital coupling\cite{OregRefaelOppen,Lutchin}.
The transport signatures of MBS are associated with anomalous $4\pi$-periodic Josephson effect
\cite{Kitaev,FuKane,1,2}. On the other hand, charge-neutral chiral Majorana modes ($\chi$MM) which are, for instance, supported by SC-magnet junctions on 3D TI surface \cite{FuKane3DTI, FuKaneMachZehnder}, are expected to show unconventional interferometry if 
involved in charge transfer between normal metal leads. Various $\chi$MM-based devices operating as  Mach-Zehnder \cite{FuKaneMachZehnder, AkhmerovMachZehnder}, Fabry-P\' erot \cite{LawFabryPerot, ButtikerFabryPerot}, and Hanbury-Brown Twiss \cite{HBT} quantum interferometers have been proposed. In context of quantum computation they can serve as a readout tool of qubit states encoded by MBS in vortex cores \cite{AkhmerovMachZehnder}.

In the paper we   calculate DC Josephson current-phase relation (C$\Phi$R) of a long spinless contact between 1D gapless  $\chi$MMs. The system under consideration represents a chiral Fabry-P\' erot interferometer implemented on a surface of 3D TI 
partially gapped by SC and  magnetic (M) islands.
Similar ideas were explored in Refs.
\onlinecite{FuKaneMachZehnder, AkhmerovMachZehnder,LawFabryPerot,ButtikerFabryPerot,HBT}.  In these works normal dissipative transport of Dirac fermions influenced by their splitting to Majorana modes and the interference of the latter were studied. In contrast, we study the equilibrium phenomenon of Josephson current carried by the Andreev states formed 
in the chiral Dirac liquid. The latter connects between two superconducting leads, which are effectively gapless 
due to chiral Majorana modes at their edges.
Spinless structure of the junction follows from the absence of spin degeneracy of 2D  helical surface states. Presence of Zeeman field breaks time-reversal symmetry and produces gapless chiral channels at the magnetic domain boundaries.  
In our system the normal conducting region consists of two separated chiral Dirac channels supported by magnetic domain walls.

We explore two models  of (i) transparent and (ii) tunnel  junctions. The first  one involves four chiral Y-junctions, operating as direct charged-to-neutral fermion converters. The second one has two tunnel contacts  formed by thin layers of magnetic material  where Majorana and normal Dirac  channels are hybridized. 
The length of the N-region, formed by two counter propagating 1D Dirac modes, is assumed to be larger than coherence length of the induced superconductivity. In other words, the Thouless energy of the normal conducting part of the system under consideration is significantly smaller  than the proximity gap.

We take into account  a contribution to the  transport from sub-gap 1D chiral states only, neglecting  2D bands. 
Since our superconducting leads are gapless due to the presence of $\chi$MMs, the 
spectral current is continuous and consists of smeared Andreev levels. Assuming that superconducting leads are large and have a fixed chemical potential, we obtain the conventional $2\pi$-periodicity of the C$\Phi$R. In other words the 
non-equilibrium $4\pi$-periodic Josephson effect (which is predicted~\cite{Jiang11} for systems with zero-energy MBSs) turns out to be irrelevant here. 

The chiral structure of the normal part consisting of two 1D spinless  Dirac modes allows to make Andreev pairs non-local. This means that the pair resides on two separated Dirac channels  and one can  apply magnetic flux $f$  inducing Aharonov-Bohm phase $\phi_{AB}=2\pi f/(h/e)=\pi f/\Phi_0$, where $\Phi_0=h/(2e)$. It follows that the critical current has a fractional $2\Phi_0$-periodicity rather than the conventional $\Phi_0$ one. The realization of the normal region by two separated chiral channels in the present setup is one of the most relevant distinctions from the junctions with 2D topological insulator where Andreev pairs exist on the same edge \cite{LMAY,Tkachov}.

The paper is organized as follows. In Section \ref{3dti} we present main ideas of $\chi$MMs realization in hybrid structures on a surface of 3D topological insulator. In Section \ref{mm-n-mm} we consider the regime of transparent Josephson contacts . We derive the $S$-matrix of the contacts by means of matching the field operators in the chiral channels. Next we calculate Josephson C$\Phi$R at arbitrary temperature and flux enclosed by Dirac modes. The Section \ref{mm-i-n-i-mm} addresses the tunnelling regime, in which Majorana and Dirac modes meet 
at the tunnel contacts formed by magnetic constrictions. In Section \ref{majorana-basis}  we   discuss  the Majorana  representation of charged-to-neutral fermion scattering in transparent  and tunnel  contacts.  Section \ref{conclusions} summarizes the results of this paper.
    The technical details are provided in Appendices \ref{s-t-app} and \ref{appendix}.

\section{Majorana modes in 3D TI hybrid structure}
\label{3dti}

\begin{figure}[h]
\includegraphics[width=0.57\linewidth]{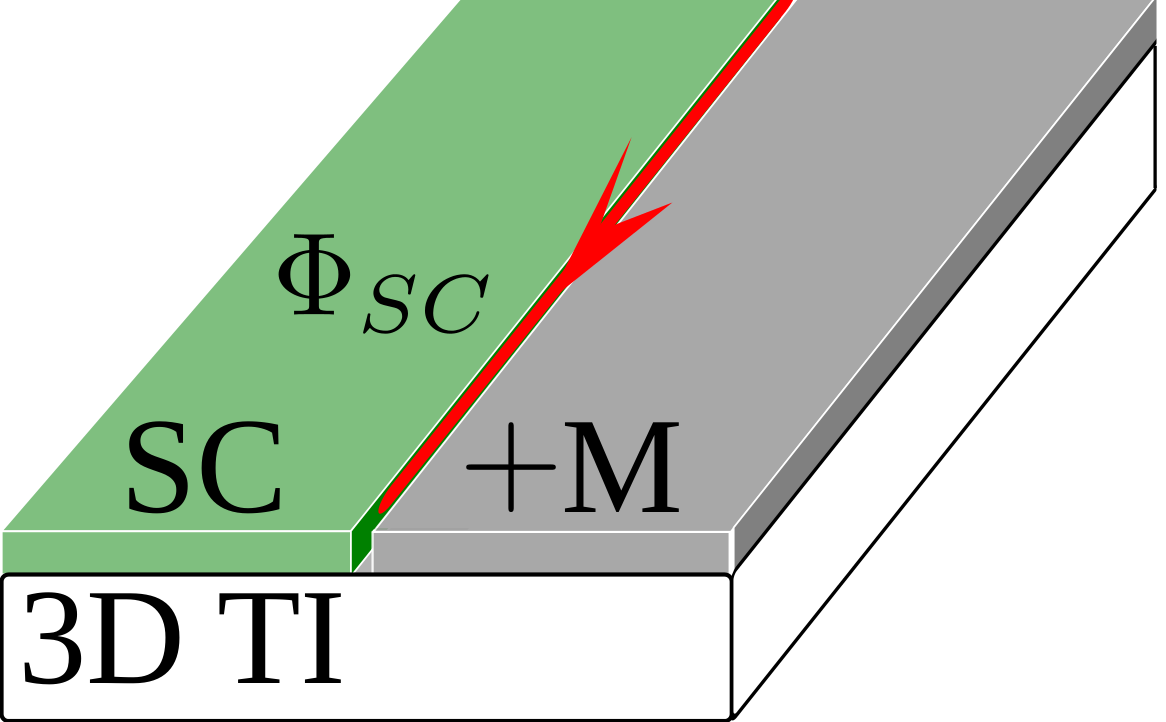}
\caption{ Superconductor/magnetic insulator boundary on the  surface of 3D topological insulator. The boundary supports a chiral Majorana mode with the chirality depending on the sign of the magnetization. } \label{chi-mm}
\end{figure}

As was proposed by Fu and Kane \cite{FuKane},  the $\chi$MM is supported by SC/magnet/3D-TI structure like that  shown in Figure \ref{chi-mm}, with an effective low-energy 1D Hamiltonian following from a solution of a 2D Bogolyubov-de Gennes equation.
The 2D surface states are described by the Rashba Hamiltonian. The system is assumed to have a chemical potential $\mu=0$, 
corresponding to the Dirac point. The induced $s$-wave SC pairing potential is given by $\Delta e^{i\Phi_{SC}}$ in the $y<0$ half plane, 
while at $y>0$ there is magnetic material inducing Zeeman exchange field $M\sigma_z$. The full Hamiltonian reads
\begin{eqnarray}
H&=&\frac{1}{2}\int dx dy \Psi^+ h \Psi\ ,\nonumber\\  h&=&iv\tau_z \mathbf{z} \cdot (\boldsymbol \sigma \times \boldsymbol{\nabla})
-M\theta(y)\sigma_z \nonumber \\&+&(\tau_+\Delta e^{-i\Phi_{SC}}+\tau_-\Delta e^{i\Phi_{SC}})\theta(-y), \label{H}
\end{eqnarray}
 where the  field  operator of the 2D surface is $\Psi = [\psi_\uparrow,\psi_\downarrow,\psi_\downarrow^+, -\psi_\uparrow^+]^T$, and 
 $\boldsymbol{\sigma}$ and $\boldsymbol{\tau}$ are Pauli matrices in spin and Nambu spaces, respectively. The field $\Psi$ satisfies  the charge conjugation constraint $\Psi=\sigma_y\tau_y \Psi^*$, while the eigenstates obey $\xi_p=\sigma_y\tau_y \xi_{-p}^*$. The low-energy $|\varepsilon_p|<\Delta,M$ eigenvalues of the Bogolyubov-de Gennes equation $h\xi_p=\varepsilon_p\xi_p$ are  single degenerate and correspond to a 1D chiral mode with the linear spectrum
 \begin{equation}
\varepsilon_p=  {\sign }(M) v p. \label{dispersion}
  \end{equation}
  Single degeneracy of eigenvalues implies that  Bogolyubov quasiparticle operator  
  \begin{equation}
  \label{eq:chip}
  \chi_p=\int dxdy (\xi_p^{(M)})^\dag(x,y)\cdot \Psi(x,y)
  \end{equation}
  represents a Majorana mode obeying $\chi_p=\chi^+_{-p}$. The
corresponding wave function of the $\chi$MM  reads
    \begin{equation}
  \xi^{(M)}_p(x,y)= \frac{1}{2}\,g^{(M)}(y)\begin{bmatrix}
e^{i(\frac{\pi}{4}-\frac{\Phi_{SC}}{2})}\\  \\
-{\sign (M)}e^{-i(\frac{\Phi_{SC}}{2}+\frac{\pi}{4})}\\ \\
-{\sign (M)}e^{i(\frac{\Phi_{SC}}{2}+\frac{\pi}{4})}\\  \\
-e^{i(\frac{\Phi_{SC}}{2}-\frac{\pi}{4})}\\  \\
\end{bmatrix} e^ {ipx}. \label{xi}
  \end{equation}
  The momentum $p$ here is directed along $x$-axis and $g^{(M)}(y)=e^{y(l^{-1}_{sc}\theta(-y)-l^{-1}_m\theta(y))}/\sqrt{2(l_{sc}+l_m)}$ is the transverse shape of 1D guiding channel. The coherence lengths (transversal decay lengths) are given by $l_{sc}=\hbar v /\Delta$ and $l_m=\hbar v /M$. The superscript $(M)$ in Eqs.~(\ref{eq:chip}) and (\ref{xi}) emphasizes the Majorana nature of the mode.

Another building block of chiral interferometers is a domain wall on a surface of a 3D TI where the magnetization sign is changed. If we consider such a   $-$M/$+$M boundary along the $x$-axis, which is described by the term $M\sigma_z {\sign}(y)$ in the Bogolyubov-de Gennes Hamiltonian,  we end up with the  Dirac chiral mode with the same spectrum   found for $\chi$MM (\ref{dispersion}). In the Nambu notation any of $\varepsilon$ eigenvalues are doubly degenerate and related to orthogonal electron and hole eigenstates $\xi_p^{(e)}$ and $\xi_{p}^{(h)}=\sigma_y\tau_y  \xi_{-p}^{(e)*}$, where
 \begin{equation}
 \xi_p^{(e)}(x,y)=\frac{1}{\sqrt {2}}\,g^{(D)}\begin{bmatrix}
 1\\  \\
 i{\sign (M)}\\ \\
 0\\  \\
 0\\  \\
 \end{bmatrix}e^ {ipx} \label{wavefunction-e}
 \end{equation}
 and $g^{(D)}=\exp(-| y|/l_m)/\sqrt{2l_m}$.
There are two independent excitations with energy $\varepsilon$ in Nambu notation in this $-$M/$+$M case   -- the electron  with momentum $p$ and  Bogolyubov operator $\psi_p$ and the $-p$ hole with $\psi_{-p}^+$. This field is complex  $\psi_p\neq\psi^+_{-p}$ and corresponds to a charged mode.
We introduce here 1D operators $\psi(x), \chi(x)$
$$
\chi(x)=\chi^+(x)=\int\frac{dp}{2\pi}\chi_p e^{-ipx}, \quad \psi(x)=\int\frac{dp}{2\pi}\psi_p e^{-ipx}
$$
with the anticommutation rules given by   $$
\{\psi^+(x_1),\psi(x_2)\}=\delta(x_1-x_2),$$
$$ \{\psi(x_1),\psi(x_2)\}=0,$$ $$\{\chi(x_1),\chi(x_2)\}=\delta(x_1-x_2)\ .$$
The same relations hold in the Heisenberg picture at equal times $t_1=t_2$.
After that the secondary quantized operators in the low energy range $|\varepsilon|<\Delta,M$ can be found as  sums over subgap chiral states. For the charge-neutral Majorana excitations we obtain
$$\Psi_M(x,y)=\int\frac{dp}{2\pi}\xi^{(M)}_p(x,y)\chi_p, $$
while for the Dirac mode the field reads
$$ \Psi_D(x,y)=\int\frac{dp}{2\pi} \left( \xi^{(e)}_p(x,y)\psi_p+\xi^{(h)}_p(x,y)\psi_{-p}^+  \right ) .$$ 
Substituting these fields into the second quantized Bogolyubov-de Gennes Hamiltonian and performing the transverse integration we obtain 
the following 1D Hamiltonians:
 \begin{eqnarray}
H_M={\sign}(M)\frac{iv}{2}\int\! dx \chi(x)\partial_x\chi(x)  \label{h-m} 
\end{eqnarray}
and
 \begin{eqnarray}
H_D={\sign}(M)iv\int\! dx \psi^+(x)\partial_x\psi(x). \label{h-d}
\end{eqnarray}

The  Hamiltonians  (\ref{h-m}) and (\ref{h-d}) correspond to  coherent propagation of the excitations in 1D guiding channels with the Fermi velocity $v$ and chirality dependent on the sign of the magnetization. The $1/2$ in the Majorana Hamiltonian $H_M$ reflects the fact that independent excitations in the $\chi$MM can be considered either at positive or  negative energies only. Say, the bottom   branch of the chiral mode (\ref{dispersion})  at $p<0$ is redundant.

The wave functions $\xi^{(M)}$ or $\xi^{(e)}$ show that the spin direction is locked to the momentum. More specifically, 
the spin textures of the guiding channels of Majorana and Dirac modes are orthogonal to the momentum direction in the particular case of Rashba type Hamiltonian (\ref{H}). 
As a consequence the spin textures are antiparallel in the counterpropagating channels. 
More generally, any scattering at a junction will be accompanied by the corresponding spin rotation.

 \section{Transparent regime}
 \label{mm-n-mm}
 \subsection{$S$-matrix of a transparent contact}
\begin{figure}[h]
\center\includegraphics[width=\linewidth]{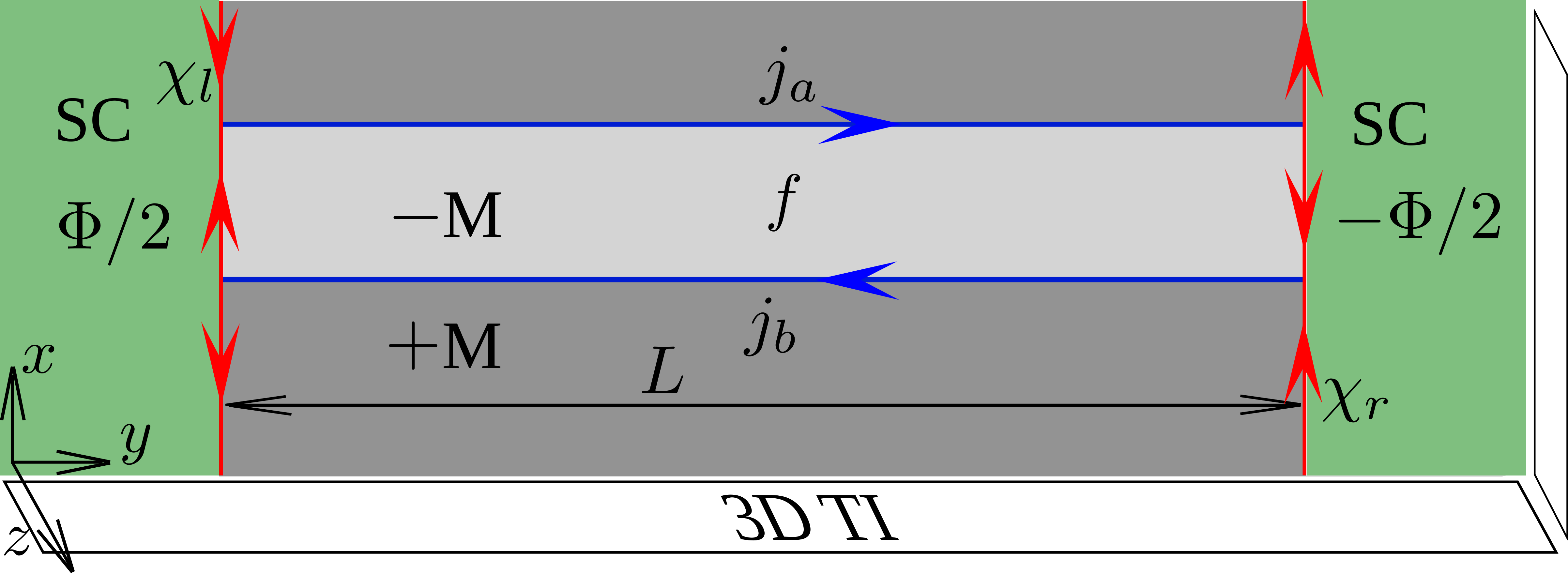}
\caption{Scheme of transparent Josephson junction on the surface of 3D topological insulator. Red  lines stand for gapless Majorana fermion  channels and arrows reflect chiralities. Superconducting electrodes, marked as green, have  phase difference $\Phi$. Light and dark grey bars are magnetic materials which induce exchange fields of the opposite polarizations $\pm$M. The line of the sign  change of M supports chiral charged modes marked as blue.   Magnetic flux $f$ in $-$M region induces   Aharonov-Bohm phase  $\phi_{AB}=\pi f/\Phi_0$.} \label{hbt}
\end{figure}

\begin{figure}[h]
\includegraphics[height=0.4\linewidth]{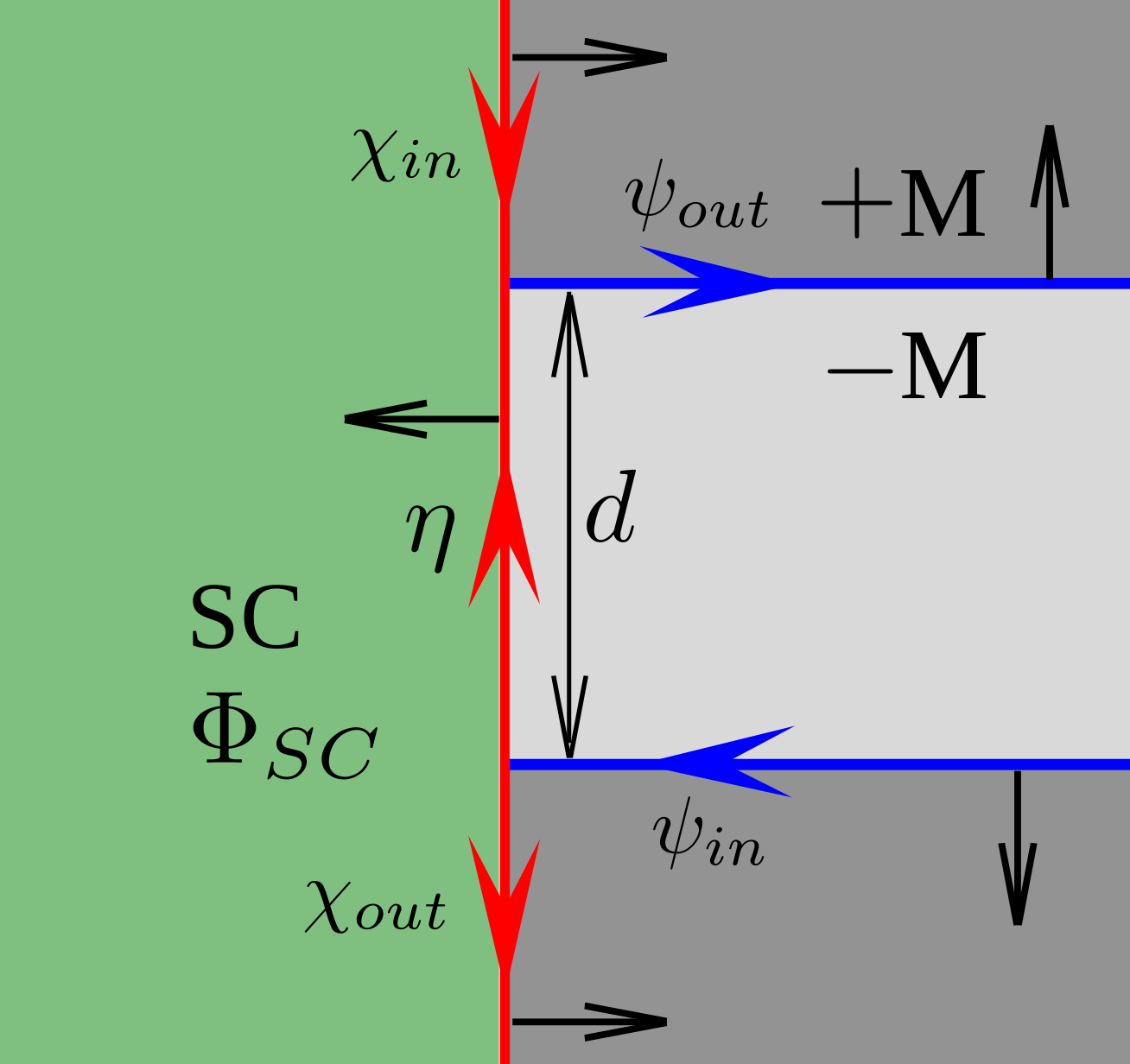}
\caption{ Structure of the contact. Black arrows shows spin texture of the chiral modes.
} \label{hbt-contact}
\end{figure}

In this section we consider the Josephson junction shown in Figure \ref{hbt}. 
This consists of two counterpropagating chiral Dirac modes which scatter at the left and right 
contacts with the superconducting leads.
Each of these contacts consists of a pair of chiral Y-junctions where electrons convert into a pair of Majorana fermions with opposite chiralities 
or  vice versa (see Figure \ref{hbt-contact}).
 The full $S$-matrix of this contact is derived from two operator relations for both upper and lower Y-junctions, described by $S_{out}$ and  $S_{in}$ matrices  found  in Refs. \onlinecite{AkhmerovMachZehnder, FuKaneMachZehnder}. Let us assume first that $\Phi_{SC}=0 $ in the electrode. The matrix $S_{in}$ involves phase difference between  electron and hole converting into two Majorana fermions. This phase is denoted  as $\alpha$ and is included in the scattering matrix as follows
 \begin{equation}\label{eq:Sin}
 \begin{bmatrix}\eta_{out} \\ \\
 \chi_{out}
 \end{bmatrix}=
  S_{in}
   \begin{bmatrix}\psi_{in} \\ \\ \psi_{in}^+
 \end{bmatrix}
  \end{equation}
   \begin{equation}
  \quad S_{in}=\begin{bmatrix}
1/\sqrt{2} &&  1/\sqrt{2}\\ \\ i/\sqrt{2}  && -i/\sqrt{2}
\end{bmatrix}\begin{bmatrix}
e^{i\alpha} &&  0\\ \\ 0  && e^{-i\alpha}
\end{bmatrix}.\label{s-in}
 \end{equation}
 The fields entering Eq.~(\ref{eq:Sin}) are Heisenberg operators at a given energy. 
The  $S_{out}$ matrix  was found in Ref. \onlinecite{AkhmerovMachZehnder} by means of time reversal transformation of $S_{in}$ which changes the  magnetization  sign $M\to -M$
        \begin{equation}
 S_{out}(M)=S_{in}^T(-M)\label{s-out}.
    \end{equation}
We assume here that both Y-junctions have identical geometries (and, in particular, $\alpha$ is the same). In this work we set the value of $\alpha$ as a free parameter which is to be found from a particular geometry of the Y-junction. The scattering in the upper Y-junction in Figure  \ref{hbt-contact} reads as
 \begin{equation}
    \begin{bmatrix}\psi_{in} \\ \\ \psi_{in}^+
 \end{bmatrix}= S_{out}
 \begin{bmatrix}\eta_{in} \\ \\
 \chi_{in}
 \end{bmatrix}.
  \end{equation}
We match Majorana operators $\eta_{in}$ and $\eta_{out}$ for the given energy $\varepsilon$ as $\eta_{in, \varepsilon}=e^{i k_\varepsilon}\eta_{out,\varepsilon}$, where the dynamic phase $k_\varepsilon=\varepsilon d/v$ is accumulated by the Majorana excitation in course of the propagation from the lower to the upper Y-junction separated by the distance $d$. The full $S(\Phi_{SC}{=}0)$-matrix of the left contact is found straightforwardly after the  exclusion of $\eta$ fields from Eqs.~(\ref{s-in}), (\ref{s-out}) and their Hermitian conjugates. 
To reinstate a non-zero SC phase of the electrode $\Phi_{SC}$ (colored as green in Figure \ref{hbt-contact}), we employ the transformation $\psi\to e^{i\Phi_{SC}/2}\psi$, yielding
\begin{equation}
S=
\begin{bmatrix}
\frac{1}{2}e^{i(k_\varepsilon+2\alpha)} & \frac{i }{\sqrt{2}}e^{i\frac{2\alpha-\Phi_{SC}}{2}} &\frac{1}{2}e^{i(k_\varepsilon-\Phi_{SC})} \\ \\
\frac{i }{\sqrt{2}}e^{i\frac{2\alpha+\Phi_{SC}}{2}} &
0&
\frac{-i }{\sqrt{2}}e^{-i\frac{2\alpha+\Phi_{SC}}{2}}\\ \\
\frac{1}{2}e^{i(k_\varepsilon+\Phi_{SC})} & \frac{-i }{\sqrt{2}}e^{i\frac{\Phi_{SC}-2\alpha}{2}}& \frac{1}{2}e^{i(k_\varepsilon-2\alpha)}
\end{bmatrix}. \label{s-matr-hbt}
\end{equation}
This unitary $S$-matrix   has particle-hole  symmetry $$S(\varepsilon)=\mathcal{Z} S^*(-\varepsilon)\mathcal{Z},\quad \mathcal{Z}=\begin{bmatrix}
0 & 0 & 1 \\
0 & 1 & 0 \\
1 & 0 & 0
\end{bmatrix} . $$
and acts on $(\psi_{in, \varepsilon}, \ \chi_{in, \varepsilon}, \ \psi^+_{in, -\varepsilon})^T$.  
It describes the partial Andreev reflection in spinless Dirac channel, which is combined with neutral Majorana excitations. The Andreev part of this process is accompanied by a Cooper pair absorption by the SC electrode.

\subsection{Current-phase relation of transparent junction}
In the following consideration  we assume that chiral Dirac channels have equal lengths, are parallel to each other and separated by the distance $d$.  Using the above approach based on $S$-matrix  (\ref{s-matr-hbt}), we find  $\psi$-operators of charged fermions  as  linear combinations of uncorrelated field operators $\chi_{l}$ and $ \chi_{r}$ of incident Majorana modes. The latter are characterized 
by the Fermi distribution function:
 \begin{equation}
   \langle\chi_{\varepsilon,i}^\dag\chi_{\varepsilon,j}^{\phantom\dag}\rangle = \frac{v^{-1} \delta_{i,j}}{1+e^{\varepsilon/T}}, \
 \chi_\varepsilon=\chi_{-\varepsilon}^\dag=\int \chi(t) e^{i\varepsilon t}dt. \label{d-function}
\end{equation}
where $v^{-1}$ is the density of states in the $\chi$MM channels. We assume  $k_{\rm B}=1$ everywhere  and recover it in final expressions.
  Using  operator relations, we calculate   chiral local densities of states $\rho_{\varepsilon}^{(a)}, \rho_{\varepsilon}^{(b)}$ and currents $j_a, j_b$.  The linear spectrum of the system (\ref{dispersion}) means that chiral current $j_i$ is proportional to charge density and, hence, the  Josephson current  $j$ is given by
 \begin{equation}
j=j_a-j_b=-ev(\langle\psi_a^+\psi_a\rangle-\langle\psi_b^+\psi_b\rangle). \label{j0}
\end{equation}
The positive direction of the current is defined from left to right SC,  bias voltage is zero and the SC phases on the electrodes are equal to  $\pm\Phi/2$, as shown in Fig.\ref{hbt}. While solving for Dirac $\psi_{\varepsilon,i}$-operators, we take into account the 
dynamic $\varphi_\varepsilon$ and the Aharonov-Bohm $\phi_{AB}$ phases 
$$
\varphi_\varepsilon=\frac{\varepsilon}{E_T}, \quad \phi_{AB}=\pi\frac{f}{\Phi_0}.
$$
The Thouless energy  $E_T$ here equals to inverse  dwell time of the interferometer (or the level spacing in the N-region)
\begin{equation}
  E_T=\frac{\hbar v}{2L+2d}.
\end{equation}
Dynamic phase $\varphi_\varepsilon$ is accumulated by $\varepsilon$-electron or $-\varepsilon$-hole enclosing the interference loop of total length $2L+2d$. Calculations of expressions for $\psi_{\varepsilon,i}$ are analogous to those  presented in Appendix \ref{appendix} for tunnel junction.

Within this $S$-matrix formalism, valid for energies in the range $|\varepsilon|<\Delta, M$, we find the following result for the 
C$\Phi$R:  
\begin{equation}\label{eq:jJepsilon}
j(\Phi,\phi_{AB})={\frac{e}{\hbar}}\int J_\varepsilon(\Phi,\phi_{AB}) n_F(\varepsilon,T)d\varepsilon\,,
\end{equation} 
where the distribution function is determined by that of the incoming $\chi$MMs defined in (\ref{d-function}). The spectral current $J_\varepsilon$  can be associated with the local densities of states in the arms $a,b$ via $J_\varepsilon={ev}( \rho_\varepsilon^{(a)}-\rho_\varepsilon^{(b)})$. The densities of states  $\rho_\varepsilon^{(a)}$, discussed in Section \ref{dos-app}, have dimension of inverse velocity, hence, $J_\varepsilon$ is dimensionless. We obtain
\begin{multline}
 J_\varepsilon(\Phi,\phi_{AB})=\\ \frac{-\sin \varphi_\varepsilon \sin\Phi }{1+\left(\frac{\cos (\phi_{AB}+4\alpha)+\cos \Phi}{2}\right)^2-(\cos (\phi_{AB}+4\alpha)+\cos \Phi)\cos\varphi_\varepsilon}.\label{s-curr-exact}
\end{multline}
This is a non-singular $2\pi$-periodic function which reflects the structure of the broadened Andreev levels. As we mentioned in the Introduction, conventional $2\pi$-periodicity with respect to $\Phi$ is related to the assumption of equilibrium (no transport voltage). This means that there are no parity effects resulting in anomalous $4\pi$ Josephson effect.

The integration in Eq.~(\ref{eq:jJepsilon}) is reduced to a sum over positive Matsubara fermionic frequencies $iT (2\pi n +\pi)$. 
We note that the dynamical phase takes the values $\varphi_\varepsilon \to iT (2\pi n +\pi)/E_T$ at these Matsubara frequencies. 
Evaluating the sum, we obtain the C$\Phi$R 
\begin{multline}
 j(\Phi, \phi_{AB})  =  4\pi  {\frac{e k_B T}{\hbar} }\sin\Phi   \\ \times\sum \limits_{n=0}^\infty \frac{1}{2e^{\pi \frac{k_BT(1+2n)}{E_T}}-\cos (\phi_{AB}+4\alpha)-\cos \Phi } . \label{cphr-hbt}
\end{multline}
This is one of the central results of this paper. 
It describes the DC Josephson current at arbitrary temperature and takes into account contributions from the subgap 1D states.

\begin{figure}[h]
 \includegraphics[width=0.7\linewidth]{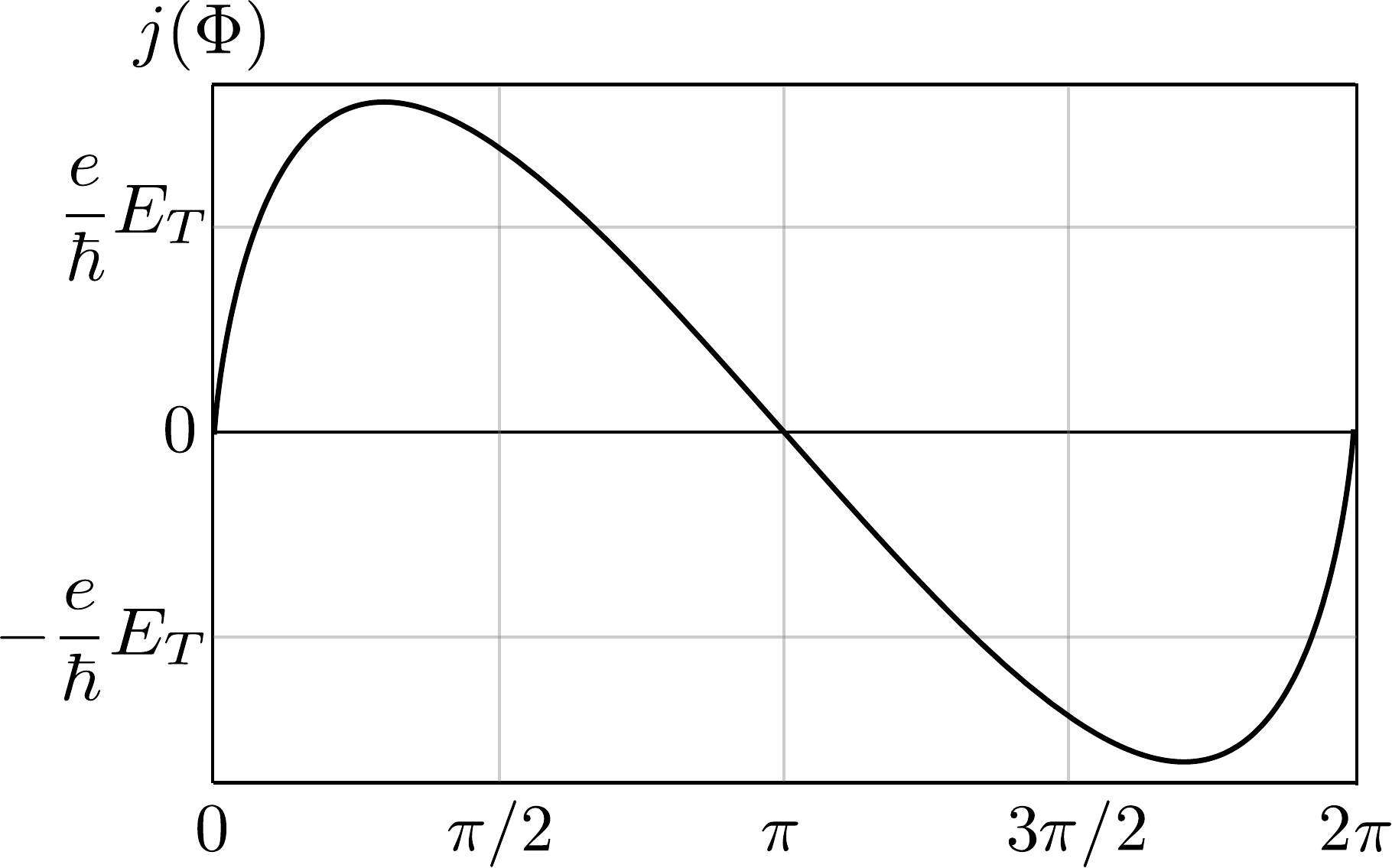}  \\
Superconducting phase, $\Phi$
\caption{$T=0$ C$\Phi$R of the transparent junction at $\phi_{AB}=\pi$. Divergent $\Phi$-derivatives at $\Phi\to 2\pi n$ are observed.} \label{cphr-transp-fig}
\end{figure}

At zero temperature the summation in $j(\Phi, \phi_{AB})$ is replaced by an integration over $x=\frac{2\pi Tn}{E_T}$ and the result reads
\begin{multline}
 j_{T=0}(\Phi, \phi_{AB})  \\ =-2  {\frac{e}{\hbar} }E_T  \frac{\ln\left( 1-\frac{\cos (\phi_{AB}+4\alpha)+\cos \Phi }{2}\right)}{\cos (\phi_{AB}+4\alpha)+\cos \Phi } \sin\Phi. \label{cphr-hbt-zerot}
\end{multline}
The  C$\Phi$R at zero temperature $T$ is presented in   Figure \ref{cphr-transp-fig}  for the value of the Aharonov-Bohm phase chosen as $\phi_{AB}=-4\alpha$. At this value the  first $\Phi$-derivative of the current is divergent as $\propto \ln|\Phi|$ at $\Phi \to 0$. This divergence  illustrates the tendency  of the  C$\Phi$R to have spikes at even  phases $\Phi=2\pi n$. In Section \ref{high-t-app} we  discuss the limit of full Andreev reflection, where the  C$\Phi$R has a sawtooth form, also with spikes at $\Phi=2\pi n$.

\begin{figure}[h]
 \includegraphics[width=0.7\linewidth]{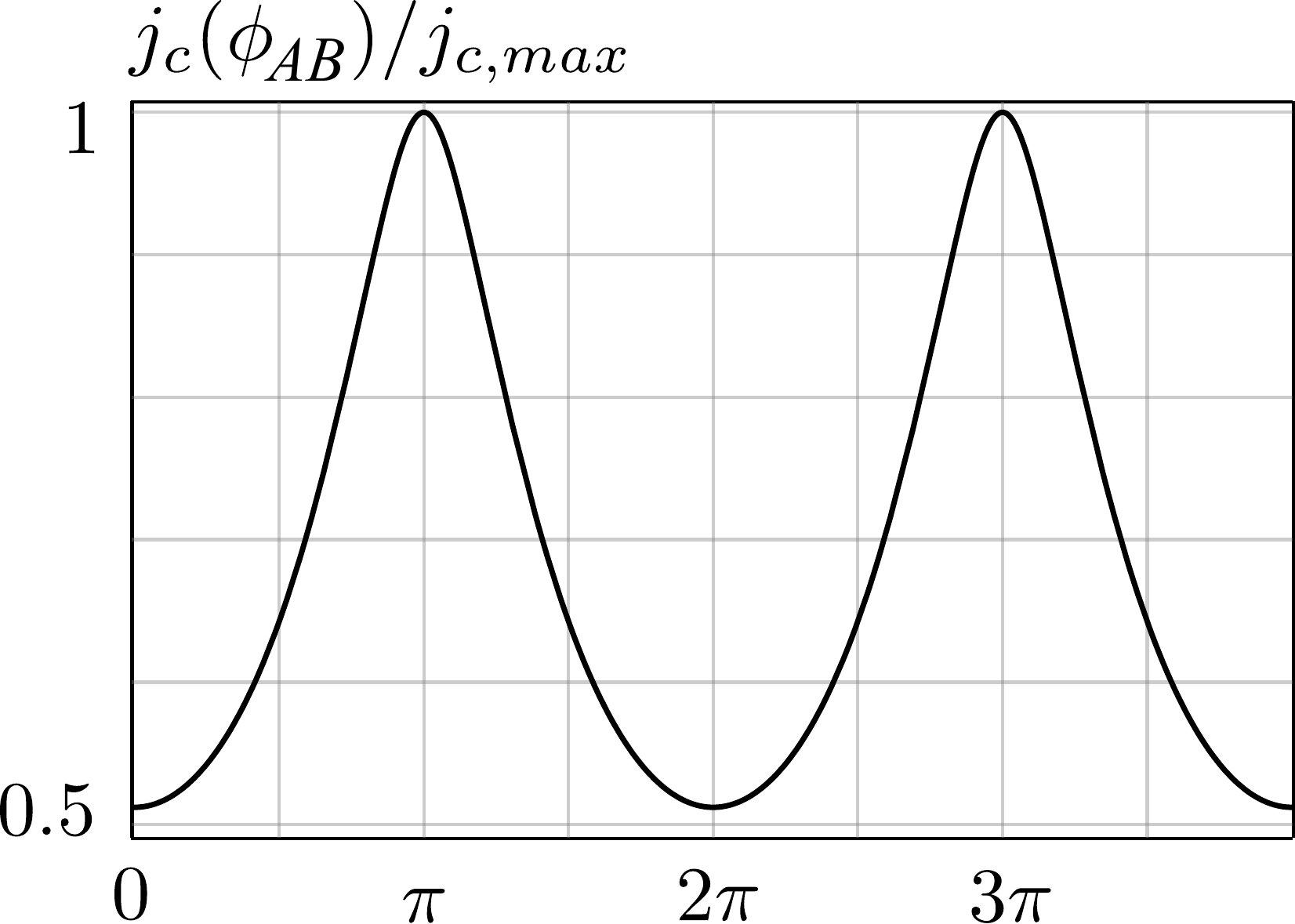}
\\
Aharonov-Bohm phase in N-region, $\phi_{AB}+4\alpha-\pi$
\caption{Fractional $h/e$-periodic pattern of critical current $j_c(\phi_{AB})$ at $T=0$. The critical current $j_c(\phi_{AB})$ is normalized to the maximal value $j_{c,max}$ and plotted as a function of  Aharonov-Bohm  phase $\phi_{AB} =\pi f/\Phi_0$, where $f$ is the magnetic flux.} \label{hbt-fig}
\end{figure}

 Two separated Dirac modes connecting the two superconductors form a SQUID loop. In view of the chirality of the junction the Andreev pair belongs to both Dirac channels, since a reflection into the same channel is forbidden. 
 Considering the junction as a SQUID loop controlled by a magnetic flux applied to the $-$M light gray bar in Fig.\ref{hbt}, we observe the fractional  $2\Phi_0=h/e$-periodic  pattern for the critical current. In Figure \ref{hbt-fig}  we plot the critical current as  a function of the flux-induced  Aharonov-Bohm phase $\phi_{AB} =\pi f/\Phi_0$. The curve is  symmetric under the assumption that $4\alpha=\pi$. (We recall that $\alpha$ is related to the geometry of the Y-junctions.) The critical current $j_c$ in this plot is normalized by its  maximal value $j_{c,max}$. At arbitrary $\alpha$ the curve in Figure  \ref{hbt-fig} would be horizontally shifted.  The positions of the peaks of $j_c$ are given by $\phi_{AB, max}= -4\alpha+2\pi n$.


\section{Tunneling regime}
\label{mm-i-n-i-mm}
\subsection{Scattering matrix of a tunnel  contact}

\begin{figure}[h]
	\center\includegraphics[width=\linewidth]{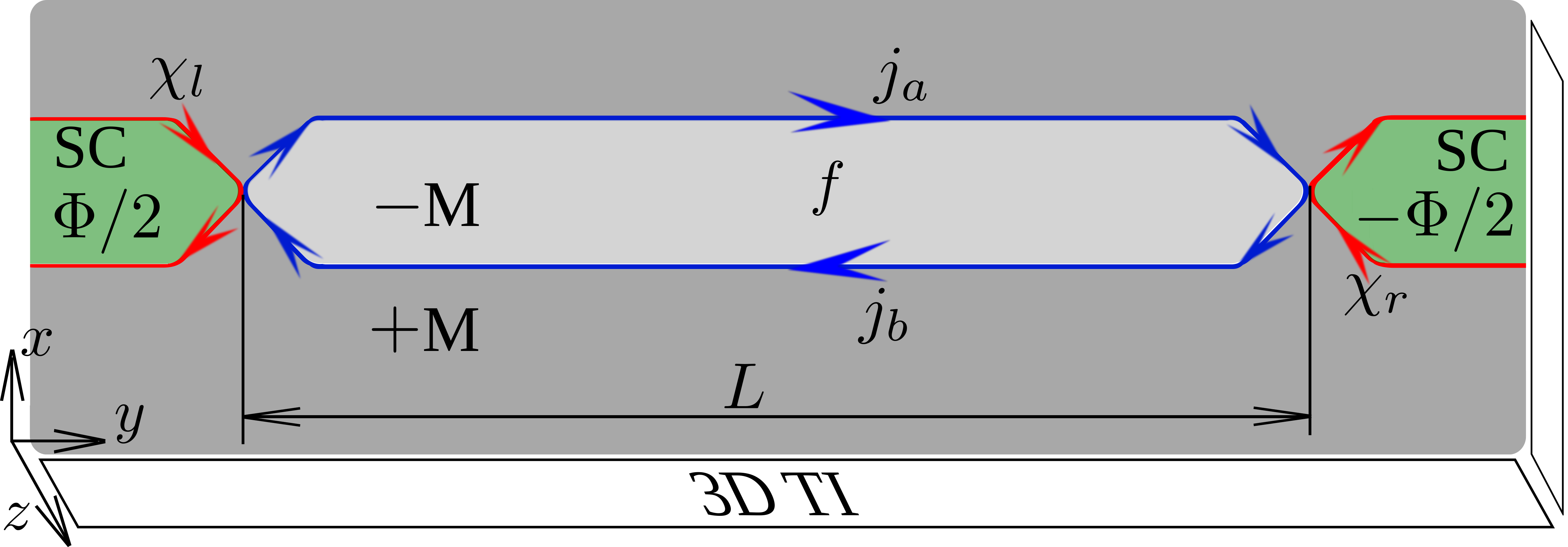}
	\caption{Scheme of the Josephson junction in the tunnelling regime. 
		 } 
		\label{setup}
\end{figure}

\begin{figure}[h]
	\includegraphics[width=0.35\linewidth]{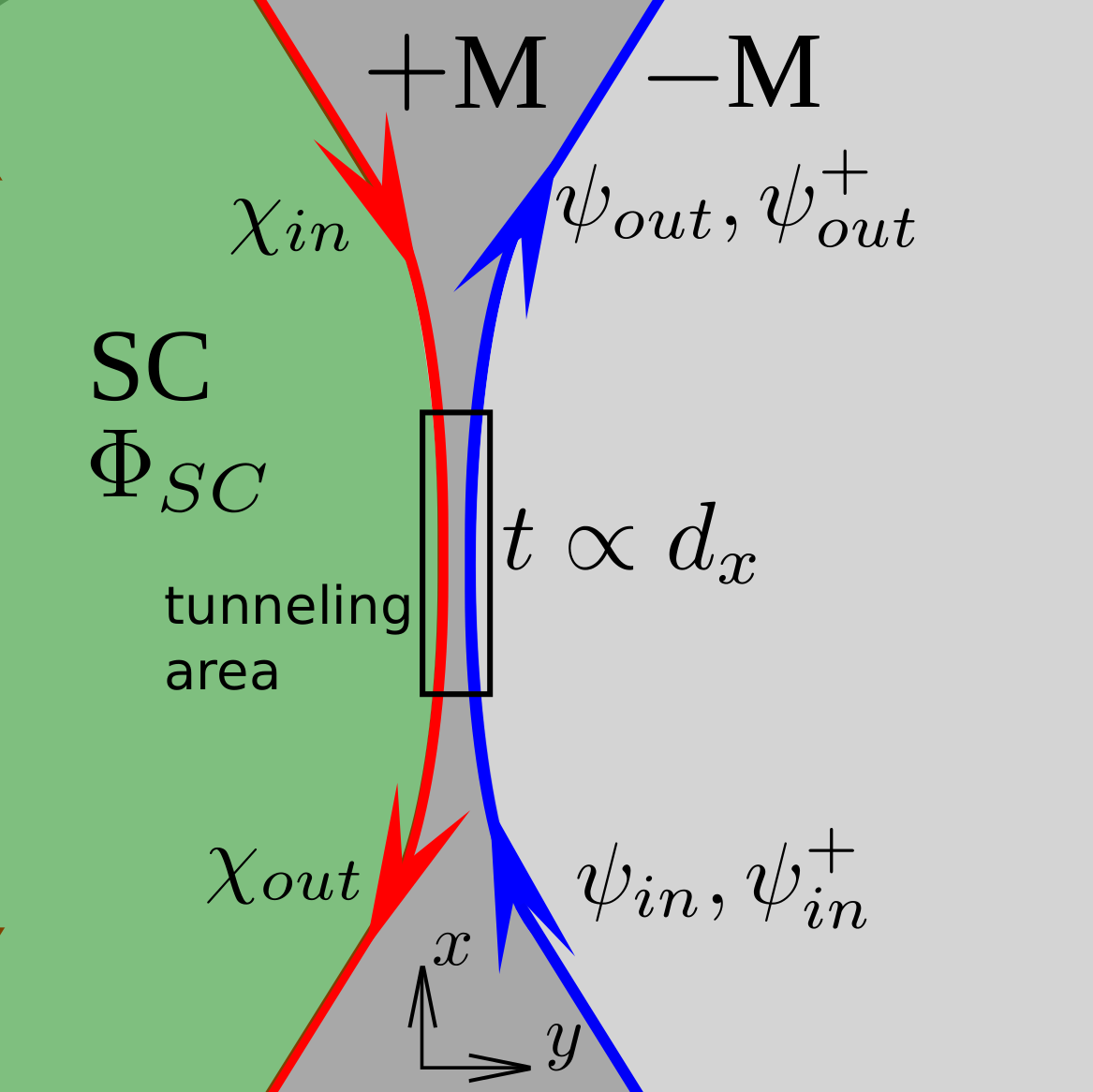}
	\caption{   Structure of the tunneling contact implemented as a constriction of the magnetic material marked as dark gray. 
	The area of hybridization between the wave functions of the counter propagating neutral and charged chiral channels is indicated by a bar.} 
	\label{contact}
\end{figure}

  The realization of the tunneling regime of the Josephson junction studied in this paper is presented in Figure \ref{setup}. 
  The left and the right tunnel contacts are implemented as constrictions of the magnetic material (see Figure \ref{contact}). Similar to the previous Section \ref{mm-n-mm}, the Dirac channels $a$ and $b$ are colored as  blue lines and  have equal lengths $l_a=l_b=L$ and geometries. In contrast to the transparent regime, in the tunneling regime the Dirac mode is not terminated but rather forms a closed loop. Hence, we introduce the total phase acquired over the loop a sum of the Berry phase $\pi$ and the Aharonov-Bohm phase, $\phi_{ext}\equiv \pi+\pi f/\Phi_0$. The constriction of the $+$M magnet plays the role of an insulating barrier.

We employ the effective 1D Hamiltonians (\ref{h-m},\ref{h-d}) and add a local  tunneling term. The resulting Hamiltonian 
reads
\begin{multline}
H= iv \int\psi^+\partial_x\psi  dx-\frac{iv}{2} \int\chi\partial_x\chi  dx +\\ +t \chi \left(\psi e^{i\frac{\Phi_{SC}}{2}}-\psi^+ e^{-i\frac{\Phi_{SC}}{2}}\right). \label{h-contact}
\end{multline}
This Hamiltonian describes counter propagating Majorana and Dirac modes which mix at $x=0$. The microscopic structure of the contact is accounted for by the small tunneling amplitude $t\ll v$, which is an additional parameter of our theory.

In Appendix \ref{s-t-app}, using Hamiltonian (\ref{h-contact}), we derive the scattering matrix $S_t$ of the tunnel 
contact acting on fields $(\psi_{in, \varepsilon}, \ \chi_{in, \varepsilon}, \ \psi^+_{in, -\varepsilon})^T$. The scattering matrix reads
\begin{equation}
S_t=
\begin{bmatrix}
\frac{1}{1+\lambda} & \frac{i\sqrt{2\lambda}}{1+\lambda}e^{-i\frac{\Phi_{SC}}{2}}&
\frac{\lambda}{1+\lambda} e^{-i\Phi_{SC}}\\ \\
\frac{i\sqrt{2\lambda}}{1+\lambda}e^{i\frac{\Phi_{SC}}{2}}&
\frac{1-\lambda}{1+\lambda}&
\frac{-i\sqrt{2\lambda}}{1+\lambda}e^{-i\frac{\Phi_{SC}}{2}}\\ \\
\frac{\lambda}{1+\lambda}e^{i\Phi_{SC}} &
\frac{-i\sqrt{2\lambda}}{1+\lambda}e^{i\frac{\Phi_{SC}}{2}}& \frac{1}{1+\lambda}
\end{bmatrix}.
\label{s-matr}
\end{equation}
The formal solution (\ref{s-matr}) obtained in Appendix \ref{s-t-app} is valid for any value of $t$ and
the dimensionless parameter $\lambda$ characterizing the scattering is given by
\begin{equation}
\lambda(t)=\tan^2\frac{t}{\sqrt{2}v} .\label{s-t-t}
\end{equation}
Yet, since the Hamiltonian (\ref{h-contact}) is physically justified only in the weak tunneling limit $t \ll v$, 
we obtain $\lambda\approx (1/2)\,t^2/v^2\ll 1$.
The amplitude of the dominant process of normal reflection is given by $1/(1+\lambda)$. 
The scattering to the Majorana channel scales as $\sim\sqrt{\lambda}$, whereas the Andreev reflection amplitude 
has the lowest amplitude $\sim \lambda$.


\subsection{Current-phase relation in the tunneling regime}
Using the scattering matrix $S_t$ calculated above (\ref{s-matr}), we obtain the following results for the spectral current 
and the C$\Phi$R (for details see Appendix \ref{appendix}):
\begin{widetext}
	\begin{equation}
	J_\varepsilon(\Phi,\phi_{AB})=\frac{-4\lambda^3 \sin{\Phi} \sin{\varphi_\varepsilon} } {\left((1+\lambda^2)\cos{\varphi_\varepsilon}+\cos{\phi_{AB}}-\lambda^2\cos{\Phi}\right)^2 + 4\lambda^2 \sin^2{\varphi_\varepsilon}}. \label{sp_current}
	\end{equation}
	\begin{equation}
	j(\Phi,\phi_{AB})=4\pi  {\frac{e k_B T}{\hbar} } \lambda^2 \sin{\Phi}\sum\limits_{n=0}^\infty \frac{1}{(1+\lambda^2)\cosh\left(\frac{\pi k_BT(2n+1)}{E_T}\right)+2\lambda\sinh\left(\frac{\pi k_BT(2n+1)}{E_T}\right)+\cos\phi_{AB}-\lambda^2\cos\Phi}. \label{jj-main}
	\end{equation}
\end{widetext}
Here the Thouless energy is given by $E_T=\hbar v/(2L)$.
From the C$\Phi$R in (\ref{jj-main}) we see that at high temperatures, $T\gg E_T$, only $n=0$ term contributes to the sum (\ref{jj-main}). In this limit we observe a sinusoidal C$\Phi$R and the critical current $j_c$ is exponentially suppressed:
 \begin{equation}
 j(\Phi) \approx \frac{4\pi e k_B T \lambda^2}{\hbar (1+\lambda)^2}\exp(-\pi k_B T/E_T) \sin{\Phi}. \label{jj-main_high_T}
 \end{equation}
The factor $\frac{\lambda^2}{(1+\lambda)^2}$ is proportional to the full transparency of the junction being the product of the 
transparencies of the left and the right contacts. This result is similar to the C$\Phi$Rs for a conventional S/N/S junction with the N-region being a long quantum wire~\cite{MaslovStoneGoldbartLoss,Kulik, S-LL-S}.
 
 In the low temperature regime, $T\ll E_T$, the C$\Phi$R is no longer sinusoidal and the critical current $j_c\sim E_T$ decays as  $1/L$. Flux dependent oscillations of $j_c$ are more sharper compared to those in the transparent regime. The resonant shape of the non-Fraunhofer $h/e$-periodic modulation is shown in Figure \ref{crit_current_fig}  for weak tunneling $\lambda=0.1$. The shift of the maximum of $j_c$ by $\pi$ results from the presence of the Berry phase.

 \subsection{Formal continuation of the tunneling solution to the regime of finite transparency}
\label{high-t-app}

The solution for the scattering matrix (\ref{s-matr}) follows formally from the tunneling Hamiltonian (\ref{h-contact})
at any value of $t$ (see Appendix \ref{s-t-app}). Yet, it is only physically justified for $t/v\ll 1$, or $\lambda\ll 1$.
We ask ourselves what happens if we extend (\ref{s-matr}), formally, to an arbitrary value of $t/v$ in the tunneling 
Hamiltonian (\ref{h-contact}) and consider the scattering matrix (\ref{s-matr}) at any value of $\lambda=\tan^2(t/\sqrt{2}v)$. 
We observe that at $\lambda=1$ the C$\Phi$R obtained for the transparent regime (\ref{cphr-hbt}) is recovered
(up to the geometry dependent phase $\alpha$). 
In other words, transparent contacts formed by Y-junctions correspond to the intermediate strength $\lambda=1$ 
of tunnel contacts, rather than to the regime $\lambda\to\infty$.

In this subsection we investigate the C$\Phi$R and the density of states in the N-region at arbitrary $\lambda$. In particular, we analyze the regime of full transparency, $\lambda\to \infty$, where the scattering matrix $S_t$ reaches the unitary limit and corresponds to the full Andreev reflection in the Dirac channel.  

Note, that the relation between $\lambda$ and $t/v$, obtained in Appendix \ref{s-t-app}, assumes a certain microscopic structure of the contacts and, thus, could also be different. We discuss this in more details in Appendix \ref{s-t-app}.

\subsubsection{Critical current and C$\Phi$Rs}
\begin{figure}[h]
 	\includegraphics[width=0.7\linewidth]{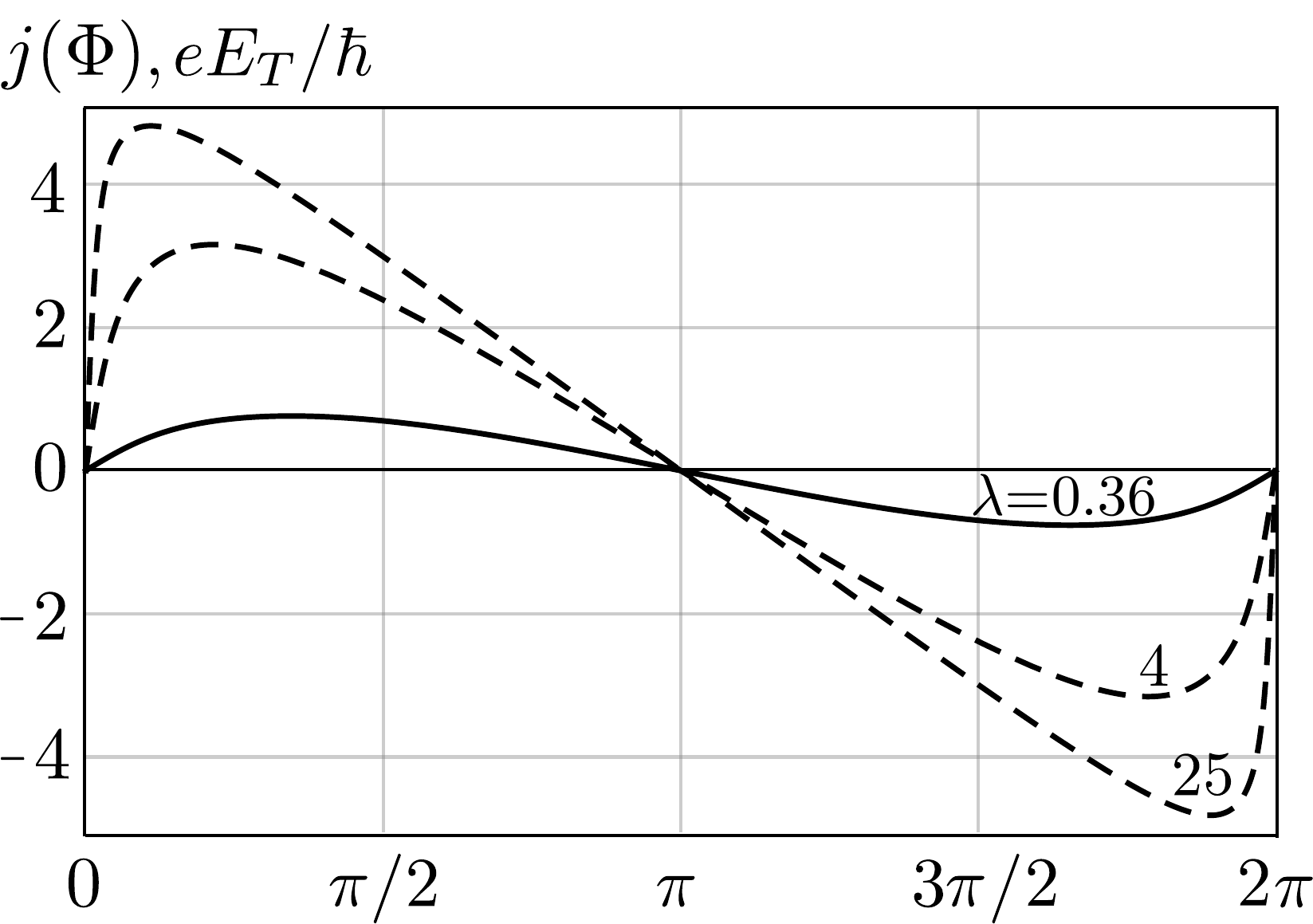}
 	\\
 	Superconducting phase, $\Phi$
 	\caption{Bold curve: C$\Phi$R found in (\ref{jj-main}) for tunnel     junction at dimensionless tunneling strengths $\lambda=0.36$. Dashed curves: continuation of the results for C$\Phi$R (\ref{jj-main}) to finite transparencies $\lambda=4$ and $\lambda=25$. The current $j(\Phi)$ is measured in units of the Thouless energy $E_T$ with Aharonov-Bohm phase $\phi_{AB}=\pi$ ($\phi_{ext}=0$) and at low temperature $T=0.01 E_T$. The $\lambda=25$ curve indicates the tendency to the formation of spikes at unusual even phases $2\pi n$.} \label{cphr_fig}
 \end{figure}

\begin{figure}[h]
 	\includegraphics[width=0.7\linewidth]{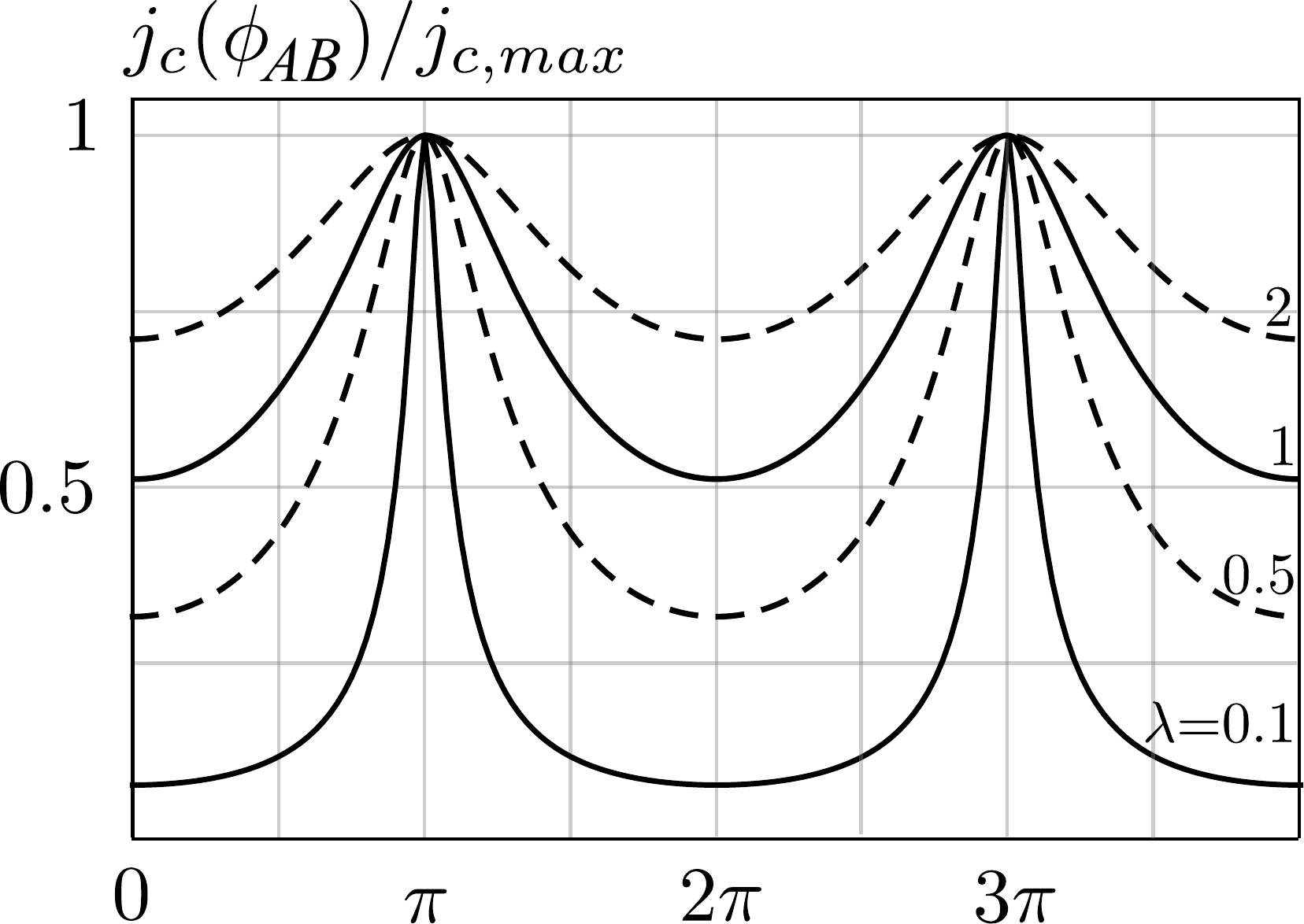}
 	\\
 	Aharonov-Bohm phase in N-region, $\phi_{AB}$
 	\caption{ Dimensionless critical current $j_c(\phi_{AB})/j_{c,max}$ of the tunnel    ($\lambda=0.1$) and transparent    at  ($\lambda=1$ and $4\alpha=\pi$) junctions  as a function of  Aharonov-Bohm  phase $\phi_{AB}=2\pi f/(h/e)$. Maximum of the critical current is shifted due to the presence of Berry phase. Dashed curves  correspond  to  results at finite   $\lambda=0.5, 2$ found from the tunneling approach. All figures are plotted at    low   temperature  $T=0.01 E_T$.} \label{crit_current_fig}
 \end{figure}

As mentioned above, the formal result for C$\Phi$R for the tunneling regime (\ref{jj-main}) is identical to that 
for the transparent regime (\ref{cphr-hbt}) at $\lambda=1$, up to geometry dependent phase $\alpha$. 
In Figures \ref{cphr_fig},\ref{crit_current_fig}  we illustrate the evolution of C$\Phi$Rs and critical currents $j_c$ upon increase of the transparency parameter $\lambda$. We plot  the results at small $\lambda=0.36$, related to tunneling approximation, and their  continuation to higher $\lambda=4$ and $25$. 
 
In the limit $\lambda\to \infty$  the scattering matrix $S_t$ (\ref{s-matr}) reaches the unitary limit
$$
 S_t(\lambda\to \infty)=	\begin{bmatrix}
 0 & 0 &
  e^{-i\Phi_{SC}}\\ \\
0 & -1 &0 \\ \\
e^{i\Phi_{SC}} & 0 & 0
 \end{bmatrix}.
 $$ 
This matrix describes the process of full Andreev reflection, where an electron converts into a hole with a phase shift of wave function equal to the phase of the SC electrode.
 At $T=0$ and $\lambda\to \infty$, where $j_c$ is independent on $\phi_{AB}$, the C$\Phi$R (\ref{jj-main}) shows a  sawtooth shape with spikes at \textit{even} phases
\begin{eqnarray}
j(\Phi)&=&4\frac{e}{\hbar}E_T \arctan\cot\frac{\Phi}{2}\nonumber\\
&=&2\frac{e}{\hbar}E_T(\pi-\Phi) \text{ for } 0<\Phi<2\pi. \label{cphr-un-lim}
\end{eqnarray}

In conventional ballistic spinful 1D S/N/S junctions the C$\Phi$Rs are known to be sawtooth-like \cite{KulikOmelyanchuk}.  
In contrast to our result  (\ref{cphr-un-lim}) the spikes are usually \textit{odd} $\pi+2\pi n$ and the slope is positive. 
This difference follows  from an additional  phase of  $\pi$ acquired by a fermion in a spinful junction after two Andreev reflections from the contacts. For spinful systems, full Andreev reflection is associated  with the time-reversal transformation $i\sigma_yK$ of the fermion wave function. Being reflected two times from the left and right contacts, the spinful fermion gets a $\pi$ phase.  In spinless junction under consideration this phase is zero.  
 
  \subsubsection{Density of states}
  \label{dos-app}
  We calculate  density of states at arbitrary $\lambda$ in $a$-branch with the use of $\psi_a$  from (\ref{psi_a}) derived in Appendix \ref{appendix}. The density of states in chiral Dirac modes are 'halfs' of the spectral current, because    $J_\varepsilon=ev(\rho_\varepsilon^{(a)}-\rho_\varepsilon^{(b)})$ if  incoming Majorana fermion modes $\chi_{r,l}$ are not correlated.   
  We introduce the local retarded Green function (calculated at an arbitrary coordinate $x$ in the $a$-branch) as $$G^{(a)}(t,t'){=}-i\theta(t{-}t')\langle \{ \psi_a(x,t),\psi_a^+(x,t')\}\rangle.$$ The Fourier transform of this function, $$G_\varepsilon^{(a)}=\int\langle\{\psi_{a,\omega},\psi^+_{a,\omega}\}\rangle \frac{ d\omega }{\varepsilon-\omega+io},$$ gives for the density of states $$\rho_\varepsilon^{(a)}=-\frac{1}{\pi} {\im} G_\varepsilon = \langle\{\psi_{a,\varepsilon},\psi^+_{a,\varepsilon}\}\rangle.$$ 
Assuming the density of states of $\chi$MM in the SC gap is constant, i.e. $\langle\{\chi_{i,-\varepsilon},\chi_{i,\varepsilon}\}\rangle=v^{-1}$, we obtain  
\begin{multline} 
 	\rho_\varepsilon^{(a)}(\Phi,\phi_{AB})=v^{-1}\\
 \frac{2\lambda\left(
 		1 + \lambda^2 +
 		\cos(\phi_{AB} -\varphi_\varepsilon) -\lambda^2  \cos(\Phi+\varphi_\varepsilon)\right) }{\left((1 +\lambda^2)  \cos\varphi_\varepsilon +
 		\cos\phi_{AB}- \lambda^2  \cos\Phi\right)^2 +
 		4 \lambda^2 \sin^2\varphi_\varepsilon} \label{dos}.
\end{multline}

\begin{figure}[h]
 	\includegraphics[width=\linewidth]{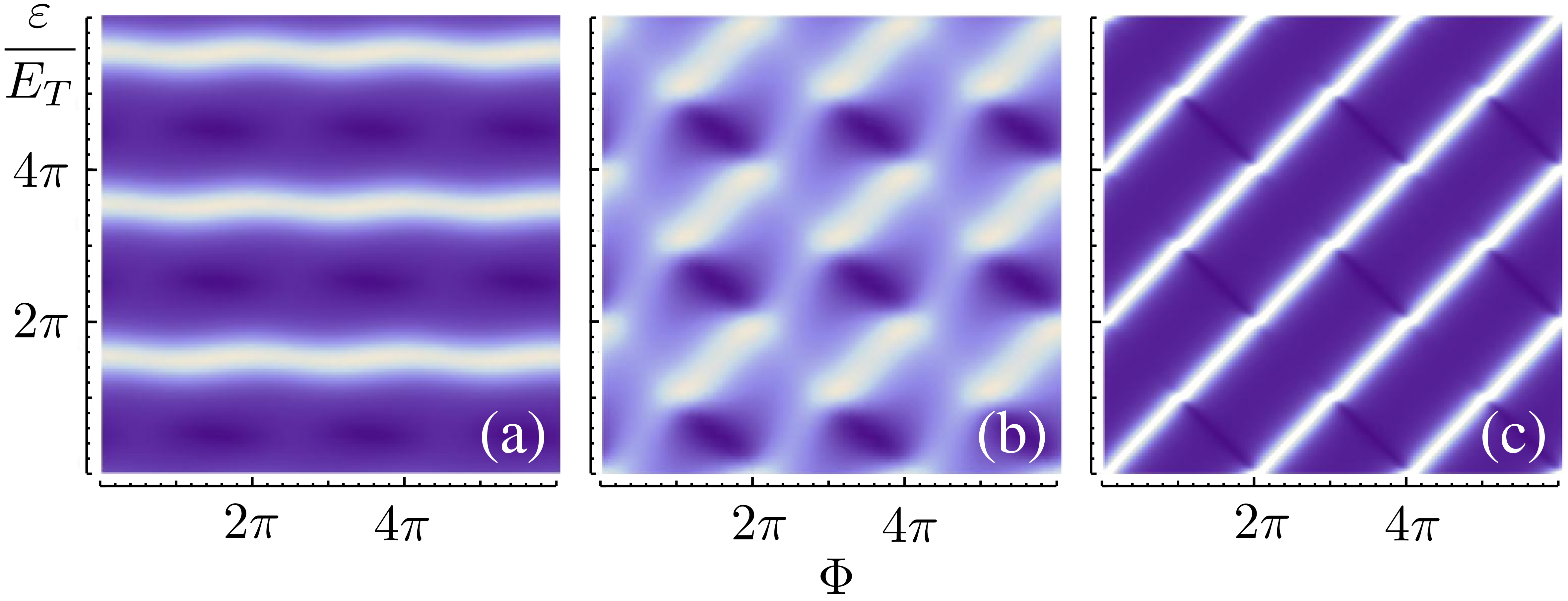}
 	\caption{Spectral density of states    (\ref{dos}) of the right movers, $\rho_\varepsilon^{(a)}(\Phi,\varepsilon)$ in the upper $a$-arm of the junction. The energy $\varepsilon$ is counted in units of Thouless energy  $E_T$. Blue and white colors correspond to low and high densities respectively. \textbf{(a)} weak tunneling regime, $\lambda=0.3$; the horizontal lines are reminiscent of the quantized levels of the isolated N-region; \textbf{(b)} intermediate transparency, $\lambda=1$, the results coincide with those for  a transparent junction at $\alpha=\pi/4$; \textbf{(c)} the case $\lambda=5$ illustrates the continuation to the high transparency limit (full Andreev reflection), where bright narrow lines are  half of full set of smeared Andreev levels.  Aharonov-Bohm phase $\phi_{AB}=-\pi/2$ for all three plots.} \label{dos_fig}
 \end{figure}

In Figure \ref{dos_fig} we plot the density of states $\rho_\varepsilon^{(a)}$ at three values of dimensionless tunneling parameter $\lambda$.  At weak tunneling, $\lambda=0.3$ (see Fig. \ref{dos_fig} (a)), the maxima of $\rho_\varepsilon^{(a)}$ are slightly dependent on $\Phi$. Horizontal lines resemble smeared mesoscopic levels of an isolated 1D Dirac wire of length $L$. These maxima can be shifted vertically by a flux induced Aharonov-Bohm phase $\phi_{AB}$, because their position is given by $$\varepsilon_n=(2\pi n- \pi-\phi_{AB})E_T.$$
The result for the intermediate transparency $\lambda=1$, equivalent to the transparent junction, is shown in Fig. \ref{dos_fig} (b). We see that the density of states is strongly smeared in this case.

 The continuation of   $\rho_\varepsilon^{(a)}$ to high transparency ($\lambda \gg 1$) shows that the spectral density is given by $2\pi$-periodic narrow lines of Andreev levels.   Their structure  can be found from the singularities of (\ref{dos}) at $\lambda\to \infty$ as
\begin{equation}
 \varepsilon_n^{A}=\pm E_T(\Phi+2\pi n). \label{and-lev}
\end{equation}
 In Figure \ref{dos_fig} (c) we plot the density of states at  $\lambda=5$ which consists of    half of the full set of smeared Andreev levels  (\ref{and-lev}). 
Note, that level positions are independent of the Aharonov-Bohm phase. This follows from the fact that the electron and 
the reflected hole get opposite Aharonov-Bohm phases, which compensate each other in closed paths.

\section{Scattering in terms of Majorana modes}
\label{majorana-basis}

One can describe  the scattering between the Dirac and Majorana chiral modes  by representing the $\psi$ operators in the N-channel with the help of two auxiliary charged-neutral Majorana operators $\gamma_1$ and $ \gamma_2$.
   These modes  do not carry  charge separately but their superpositions do.
   The  different SC phases of the opposite contacts result in a fusion of the auxiliary $\gamma_1$ and $\gamma_2$ modes, which is responsible for a Cooper pair transfer from one lead to the other.
   
   In this section we discuss the scattering  in the Majorana basis for the tunneling and transparent  contacts. 
   We analyze in detail the regime $\lambda=1$ (keeping $\alpha$ as a free parameter), since it corresponds 
   to the case of the transparent junction (\ref{cphr-hbt}) up to geometry dependent $\alpha$ and the Berry phases.

\begin{figure}[h]
\includegraphics[height=0.4\linewidth]{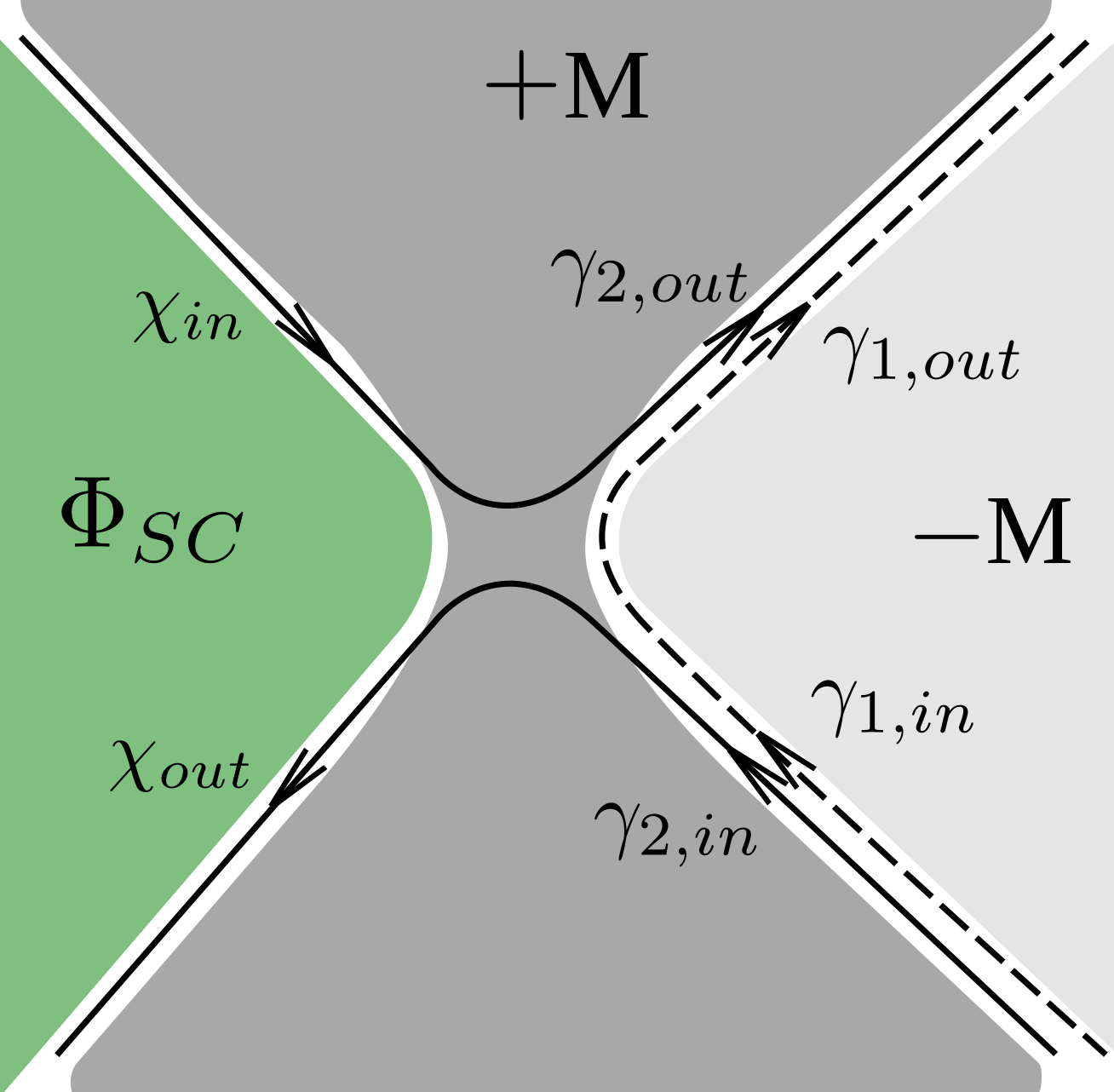}\quad
 \includegraphics[height=0.4\linewidth]{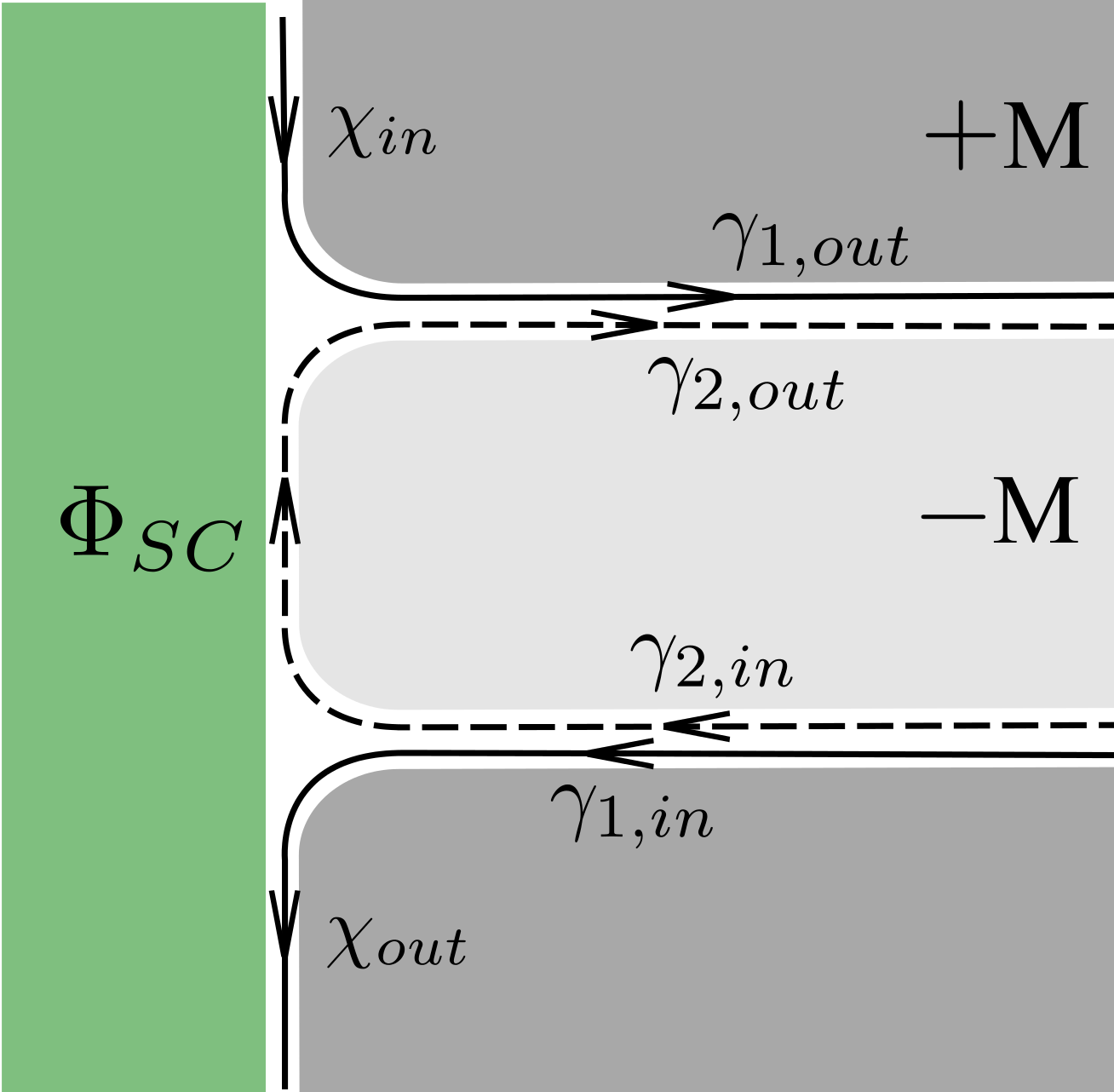}
\caption{Scattering in normal Dirac $\psi,\psi^+$ and neutral-charge channels $\chi$ written in terms of auxiliary Majorana modes $\gamma_1, \gamma_2$. Left:  tunnel     contact described by $S_t$-matrix (\ref{s-matr}) at $\lambda=1$. Right: transparent contact 
described by $S$-matrix (\ref{s-matr-hbt}) at $\alpha=\pm\pi/2$.} \label{scattering-mf}
\end{figure}

  The Majorana basis  is defined through the following  transformation for both the left and the right contacts
 \begin{equation}
   \psi_i=\frac{1}{\sqrt{2}}(\gamma_{1,i}+ i \gamma_{2,i})e^{-i\Phi_{SC}/2}, \label{basis-0}
  \end{equation}
  where the index $i$ here stands for $in$ and $out$ channels.
       The  $S_t$-matrix (\ref{s-matr}) under this transformation at arbitrary $\lambda$  reads
  \begin{equation}
\begin{bmatrix}
\gamma_{1, out } \\ \\
\chi_{out}\\ \\
\gamma_{2, out}
\end{bmatrix}=
\begin{bmatrix}
1 && 0 && 0 \\ \\
0 && \frac{1-\lambda}{1+\lambda} && -\frac{2\sqrt{\lambda}}{1+\lambda} \\ \\
0 && \frac{2\sqrt{\lambda}}{1+\lambda} && \frac{1-\lambda}{1+\lambda}
\end{bmatrix}
\begin{bmatrix}
\gamma_{1, in } \\ \\
\chi_{in}\\ \\
\gamma_{2, in }.
\end{bmatrix}\label{m-basis-tunn}
\end{equation}
According to (\ref{m-basis-tunn}) $\gamma_1$ mode is always fully decoupled, consistent with the 
scattering theory of Li, Fleury and B\"uttiker \cite{ButtikerFabryPerot}. In the weak tunneling limit $\lambda\ll 1$ 
the modes  $\chi$ and $\gamma_2$ scatter into each other with the amplitude $\sim\sqrt{\lambda}$. 
In the opposite  unitary limit $\lambda\to \infty$ all modes are decoupled from each other but both $\chi$ and $\gamma_2$ 
invert signs meaning that the corresponding Dirac fermions experience the full Andreev reflection. 
In the intermediate case $\lambda=1$ (Figure \ref{scattering-mf}, left) the modes  $\chi$ and $\gamma_2$ 
fully convert into each other:   $\gamma_{2,out}=\chi_{in}$ and $\chi_{out}=- \gamma_{2,in}$.

The presence of the scattering phase $\alpha$ in the scattering matrix (\ref{s-matr-hbt}) of the  transparent 
Dirac-Majorana contacts changes the situations considerably. 
We apply again the transformation (\ref{basis-0}) to $\psi$-operators and obtain the following $S$-matrix  
in the Majorana basis (setting $d=0$) 
\begin{equation}
\begin{bmatrix}
\gamma_{1, out } \\ \\
\chi_{out}\\ \\
\gamma_{2, out}
\end{bmatrix}=
\begin{bmatrix}
\cos^2\alpha && -\sin\alpha && -\frac{\sin2\alpha}{2} \\ \\
-\sin\alpha && 0 && -\cos\alpha \\ \\
 \frac{\sin2\alpha}{2} && \cos\alpha && -\sin^2\alpha
\end{bmatrix}
\begin{bmatrix}
\gamma_{1, in } \\ \\
\chi_{in}\\ \\
\gamma_{2, in }
\end{bmatrix}.\label{m-basis-exact}
\end{equation}
This matrix coincides with that of (\ref{m-basis-tunn}) in the limit $\lambda =1$ only if $\alpha=0$. 
For the  other values of $\alpha $ all the modes are mixing with each other. At arbitrary $\alpha$ there is  mixing between all of the modes  except of $\chi_{in}$ and $\chi_{out}$. Mixing between $\chi_{in}$ and $\chi_{out}$ is possible if we a add second contact with a different SC phase.  We illustrate the scattering in the particular case of $\alpha=\pm\pi/2$ in Figure \ref{scattering-mf} (right), where the incoming $\chi$-mode converts into $\gamma_1$ mode and vise versa. Note that, in this case, $\gamma_2$ mode is not converted to the others but gets an inversion of the sign, $\gamma_{2, out}=-\gamma_{2, in}$. 

 \section{Summary}
 \label{conclusions}
To conclude, we analyzed two limits of 1D long ballistic Josephson junctions where the leads are formed by gapless 1D chiral Majorana channels.These junctions can be realized  as hybrid structures involving 3D topological insulator surface in proximity with s-wave superconducting electrodes and magnetic materials.  The normal region of these setups is formed by two chiral Dirac modes spaced by a magnetic material. Such a normal part is effectively spinless because the spin textures are locked to profiles of magnetic domain walls. 

In the first part of the work we have calculated the C$\Phi$R in the regime of high transparency. In this limit the left and the right contacts of the junction consist each of two Dirac-Majorana converters built by magnet/superconductor interfaces. We find that this system has a continuous spectral current, a $2\pi$-periodic non-sinusoidal C$\Phi$R at low temperatures and  an $h/e$-periodic dependence of the critical current on the magnetic flux. The critical current amplitude at zero temperature is given by Thouless energy  which is proportional  to inverse dwell time of the normal region. The junction can act as a SQUID because the two Dirac channels are spatially separated. An Andreev pair in this case appears to be spatially non-local, which offers a possibility of inducing an internal Aharonov-Bohm phase. This leads to one of our central results: the critical current shows fractional $h/e$-periodic pattern. 

In the second part of the work we have studied another realization of the junction where Dirac and Majorana channels are coupled through a tunnel barrier. The C$\Phi$R of such a tunnel junction was found in terms of the  tunneling Hamiltonian approach and corresponding scattering matrix. The resulting critical current patterns show sharper resonant peaks as compared to those for the transparent junction. We have also studied the formal extension of the tunneling solution to the high transparency regime and compared it with the one obtained in the first part of the work.

Interferometers involving chiral Majorana modes \cite{FuKaneMachZehnder, AkhmerovMachZehnder,LawFabryPerot,ButtikerFabryPerot,HBT} could find their applications as measuring devices of topological qubits~\cite{AkhmerovMachZehnder}. In this paper we have explored a dual setup, in which the equilibrium Josephson current is carried by interfering chiral Dirac electrons between chiral Majorana leads. Embedded into schemes 
with vortices and/or magnetic or SC islands supporting zero-energy Majorana pairs, a Josephson based quantum interferometer might be of advantage, e.g., for performing quantum readout of topological qubits.

 \section{Acknowledgments}
 The authors thank  I.V. Protopopov, I.S. Burmistrov, A.L. Rakhmanov, A.V. Rozhkov,  P.M. Ostrovsky, W.V. Pogosov,  S.N. Artemenko and, especially, Yu. Makhlin for fruitful discussions. We acknowledge financial support by DFG Priority Program 1666, by German-Israeli Foundation, and by the EU Network Grant InterNoM. The research of A.D.M. was supported by the Russian Science Foundation under grant No. 14-22-00281.

\appendix

\section{Derivation of the scattering matrix of a tunnel contact}
\label{s-t-app}

The local form of tunneling Hamiltonians (\ref{h-contact}) is an approximation, because, microscopically, such contacts are formed by constrictions
of finite size as depicted in Figure \ref{contact}. If the constriction is wide and the hybridization 
is negligible, the parallel chiral channels along SC/+M and +M/$-$M junctions
have opposite (orthogonal) spin textures. In the tunneling area, shown as the constriction of the magnetic material in the black bar, the eigenfunctions can be approximated by superpositions of the eigenfunctions of the isolated Dirac and Majorana channels, $\xi^{(e)},\xi^{(h)}$ and $\xi^{(M)}$. 
This means that an electron coming into the tunneling area starts to oscillate between hole- and Majorana-like states with a rate, proportional to a hybridization of the channels, estimated as $Me^{-Md_y}$, where $d_y$ is a transversal constriction length.  Finally, the incoming  electron scatters into a superposition of outgoing electron, hole  and neutral excitation in the Majorana channel. 
From this qualitative picture one can conclude that the scattering matrix should be periodic with respect to the phase of these oscillations, given as a product of hybridization energy and the constriction dwell time $Me^{-Md_y/v} d_x/v$. 

Here, we derive this periodic behavior of the $S_t$-matrix from the local tunneling Hamiltonian  (\ref{h-contact}).  We employ the Heisenberg equations of motion
\begin{equation}
\begin{bmatrix}
(v\partial_x-\partial_t)\psi(x)\\ \\
(v\partial_x-\partial_t)\chi(-x)\\ \\
(v\partial_x-\partial_t)\psi^+(x)
\end{bmatrix}
=it\delta(x)
\mathcal{T}
\begin{bmatrix}
\psi(x)\\ \\
\chi(x)\\ \\
\psi^+(x)
\end{bmatrix},
\label{eq2dagger}
\end{equation}
where the tunnel matrix is given by
\begin{equation}
\mathcal{T}=\begin{bmatrix}
0 & e^{-i\frac{\Phi_{SC}}{2}} & 0 \\ \\
e^{i\frac{\Phi_{SC}}{2}}& 0 & -e^{-i\frac{\Phi_{SC}}{2}}\\ \\
0 & -e^{i\frac{\Phi_{SC}}{2}}& 0
\end{bmatrix}, \label{t-matr}
\end{equation}
We write $\chi(-x)$ in these equations  in order to make  the chirality of the Majorana mode the same as that of the charged channel. In this representation we can consider incoming and outgoing states as those at $x<0$  and at $x>0$ respectively.

It follows from the $x$-integration of (\ref{eq2dagger},\ref{t-matr}) around the point of contact $x= 0$ that the relation between the tunneling matrix $\mathcal{T}$- and the scattering matrix  $S_t$ reads
\begin{equation}
S_t=\exp\left[i(t/v) \mathcal{T}\right].\label{s-t-exp}
\end{equation}
Calculating the exponent we obtain
\begin{equation}
S_t=
\begin{bmatrix}
\frac{1}{1+\lambda} & \frac{i\sqrt{2\lambda}}{1+\lambda}e^{-i\frac{\Phi_{SC}}{2}}&
\frac{\lambda}{1+\lambda} e^{-i\Phi_{SC}}\\ \\
\frac{i\sqrt{2\lambda}}{1+\lambda}e^{i\frac{\Phi_{SC}}{2}}&
\frac{1-\lambda}{1+\lambda}&
\frac{-i\sqrt{2\lambda}}{1+\lambda}e^{-i\frac{\Phi_{SC}}{2}}\\ \\
\frac{\lambda}{1+\lambda}e^{i\Phi_{SC}} &
\frac{-i\sqrt{2\lambda}}{1+\lambda}e^{i\frac{\Phi_{SC}}{2}}& \frac{1}{1+\lambda}
\end{bmatrix},
\label{s-matr-app}
\end{equation}
where the dimensionless tunneling strength $\lambda(t)$ is  given by  
\begin{equation}
\lambda(t)=\tan^2\frac{t}{\sqrt{2}v} .\label{s-t-t}
\end{equation}
This tunneling $S_t$-matrix is unitary and obeys the particle-hole symmetry  $S_t(\varepsilon)=\mathcal{Z} S_t^*(-\varepsilon)\mathcal{Z}$ like $S$.  The eigenvalues of (\ref{s-matr-app}) are given by $e^{i\sqrt{2}t/v};e^{-i\sqrt{2}t/v};1$.

The solution method leading to (\ref{s-matr-app}) and (\ref{s-t-t}) is not universally accepted. Rather, a different ansatz was used in various problems on transport in 1D systems such as tunneling between edge states of QHE \cite{ChamonFreedWen}, impurity scattering in Luttinger liquid at $g=1/2$ \cite{EggerGrabert} or resonant Andreev reflection from zero-mode Majorana bound state \cite{Alicea1}. Following these works
we should have taken the local $\psi$- and $\chi$-operators at point $x=0$ as
\begin{equation}
\psi=\frac{\psi(-0)+\psi(+0)}{2}\quad,\quad  \chi=\frac{\chi(-0)+\chi(+0)}{2}\ . \label{psi-chi-1/2}
\end{equation}
Relations (\ref{psi-chi-1/2}) produce then the same solution for the scattering matrix as in (\ref{s-matr-app}). However, the parameter 
$\lambda$ is now different and is given by
\begin{equation}\label{eq:lambdatilde}
\tilde\lambda(t)=\frac{t^2}{2v^2}.
\end{equation}
The two solutions coincide in the weak coupling limit $t/v\ll 1$. For larger values of $t$ the difference 
is substantial. For instance, the unitary limit $\tilde\lambda\to\infty$ is achieved with (\ref{psi-chi-1/2}) 
and (\ref{eq:lambdatilde}) at $t\to\infty$. In contrast, with (\ref{s-t-exp}) it is reached at $t_n=\sqrt{2}(\pi/2+\pi n)v$.

We conjecture that the solution method, leading to (\ref{s-t-exp}) applies if the constriction is smooth enough, so that the 
validity of the low energy description provided by (\ref{H}) and (\ref{h-contact}) is not violated in any point of the constriction.
Then, Eq.~(\ref{eq2dagger}) is solved as a regular differential equation. On the other hand, the ansatz (\ref{psi-chi-1/2}) is probably applicable for sharp enough constrictions. This ambiguity should be resolved by solving 2D BdG equations for the constriction geometry.

\section{Derivation of C$\Phi$R in the tunneling regime}
\label{appendix}
In this appendix we present technical details of the calculation of the Josephson current. We find the C$\Phi$R  as the difference between the chiral currents in $a$ and $b$ arms by using the tunnel junction scattering matrix $S_t$, given by (\ref{s-matr-app}). We take the 1-st and the 3-d lines of  (\ref{s-matr-app}), disregarding  the outgoing Majorana field $\chi_{out}$ (line 2), and  obtain
\begin{multline}
\hspace{-0.3cm}\begin{bmatrix}
\psi_{out, \varepsilon}\\ \\
\psi_{out, -\varepsilon}^+
\end{bmatrix}
=\\
=\begin{bmatrix}
 \frac{1}{1+\lambda} & \frac{\lambda e^{-i\Phi_{SC}}}{1+\lambda} \\ \\
 \frac{\lambda e^{i\Phi_{SC}}}{1+\lambda} & \frac{1}{1+\lambda}
\end{bmatrix}
\begin{bmatrix}
\psi_{in, \varepsilon}\\ \\
\psi_{in, -\varepsilon}^+
\end{bmatrix} +
\frac{i\sqrt{2\lambda}}{1+\lambda}\chi_{in,\varepsilon}\begin{bmatrix}
e^{-i\frac{\Phi_{SC}}{2}}\\ \\
-e^{i\frac{\Phi_{SC}}{2}}
\end{bmatrix} .\label{m-norm-andr}
\end{multline}
The matrix in the r.h.s. of Eq.~(\ref{m-norm-andr}) contains  amplitudes of  normal and Andreev reflection, while the last term describes coupling with the equilibrium Majorana $in$-channel. This term is responsible for the spectral 
current (\ref{sp_current}) being continuous, due to the gapless spectrum of the incoming lead Majorana mode.

To proceed we rewrite Eq.~(\ref{m-norm-andr}) as follows
\begin{equation}
\hspace{-0.3cm}\begin{bmatrix}
\psi_{out, \varepsilon}\\ \\
\psi_{in, -\varepsilon}^+
\end{bmatrix}
=
\mathcal{M}_{\Phi_{SC}}
\begin{bmatrix}
\psi_{out,- \varepsilon}^+\\ \\
\psi_{in, \varepsilon}
\end{bmatrix} +
i\sqrt{2 \lambda}\chi_{in,\varepsilon}\begin{bmatrix}
e^{i\frac{\Phi_{SC}}{2}}\\ \\
e^{-i\frac{\Phi_{SC}}{2}}
\end{bmatrix}. \label{s-0}
\end{equation}
The matrix $\mathcal{M}_{\Phi_{SC}}$ in (\ref{s-0}) is given by
$$
\mathcal{M}_{\Phi_{SC}}=\begin{bmatrix}
\lambda e^{i\Phi_{SC}} && 1-\lambda \\ \\
 1+\lambda &&
-\lambda e^{-i\Phi_{SC}}
\end{bmatrix}.
$$

Using (\ref{s-0}) we formulate the boundary conditions for the left and right  contacts.  We introduce the field $ \Psi_{\varepsilon,x}=[\psi_{a, \varepsilon},\psi_{b, -\varepsilon}^+, \psi_{a, -\varepsilon}^+,\psi_{b, \varepsilon}]_{x}^T$, which depends 
on $\varepsilon$ and on the coordinate $x$ along the Dirac channels. Indices $a$ and $b$ stand for the upper and lower Dirac modes (see Figure \ref{setup}). We start from the left contact where $\Phi_{SC}=\Phi/2$ and $x=-L/2$.  
Using (\ref{s-0}) and its Hermitian conjugate at $\varepsilon\to-\varepsilon$ as well as the property 
of the real Majorana field, $\chi^+_{l,-\varepsilon}=\chi_{l,\varepsilon}$, we derive 
\begin{equation}
\begin{bmatrix}
\gamma_0 && -\mathcal{M}_{\Phi/2}\\ \\
-\mathcal{M}_{\Phi/2}^* && \gamma_0
\end{bmatrix}
  \Psi_{\varepsilon,-L/2}=
i\sqrt{2 \lambda}
\eta_{\Phi/2}\chi_{l,\varepsilon}.\label{bc_left}
\end{equation}
Here $\gamma_0$ is the $2\times 2$ identity matrix in the left upper block and $\eta_{\Phi/2}=[e^{i\frac{\Phi}{4}},e^{-i\frac{\Phi}{4}},-e^{-i\frac{\Phi}{4}},-e^{i\frac{\Phi}{4}}]^T$. Note, that for the left contact $b$-components of $\Psi_{\varepsilon,-L/2}$ are $in$-fields while $a$-components are $out$-fields: $\psi_{a,-L/2}=\psi_{out}$ and $\psi_{b,-L/2}=\psi_{in}$. 
The rank of the 4$\times$4 matrix in (\ref{bc_left}) equals $2$ and the eigenvalues are equal to $0,0,2,2$ regardless of the values of $\lambda$ and $\Phi$.
The condition for the right contact is obtained from (\ref{bc_left}) by the replacements
\begin{equation}
\Phi\to -\Phi, \quad \chi_l \to\chi_r, \quad \Psi_{\varepsilon,-L/2}\to s_x\gamma_x\Psi_{\varepsilon,L/2}. \label{bc_right}
\end{equation}
The Pauli matrix $s_x$, introduced in (\ref{bc_right}), mixes the $2\times 2$ blocks in (\ref{bc_left}) acting as a particle-hole transformation, while $\gamma_x$ acts inside of the blocks. The product $s_x\gamma_x$ interchanges $a$ and $b$ indices in $\Psi_{\varepsilon, x}$.
We find that the fields in the middle of N-region, $\Psi_{\varepsilon,0}\equiv \Psi_{\varepsilon,x=0}$, can be expressed via the operators 
at the ends of the Dirac channels, $\Psi_{\varepsilon,-L/2}$ and $\Psi_{\varepsilon,L/2}$, using the dynamical and geometrical phases:
\begin{eqnarray}
\Psi_{\varepsilon,-L/2}=\mathcal{D}_\varepsilon \mathcal{F}\Psi_{\varepsilon,0},  \label{x01} \\
 \Psi_{\varepsilon,L/2}=s_x\gamma_x\mathcal{D}_\varepsilon \mathcal{F}s_x\gamma_x\Psi_{\varepsilon,0}. \label{x02}
\end{eqnarray}
Here $\mathcal{D}_\varepsilon={\diag}[e^{-i\varphi_\varepsilon/4}, e^{i\varphi_\varepsilon/4}, e^{-i\varphi_\varepsilon/4}, e^{i\varphi_\varepsilon/4}]$
and $\mathcal{F}={\diag}[e^{-i\phi_{ext}/4}, e^{-i\phi_{ext}/4}, e^{i\phi_{ext}/4}, e^{i\phi_{ext}/4}]$. The  external phase  $\phi_{ext}$ equals to the sum of  Aharonov-Bohm and Berry phases and the dynamical  phase $\varphi_\varepsilon$ accumulated an electron 
of energy $\varepsilon$
or a hole of energy $-\varepsilon$, enclosing the interference loop of the length $2L$. Here
\begin{equation}
 \varphi_\varepsilon=\frac{ \varepsilon}{E_T}, \ E_T=\frac{\hbar v}{2L}.
\end{equation}
 The relations (\ref{bc_left}, \ref{bc_right}) together with (\ref{x01}, \ref{x02}) make the problem  of finding the four components of $\Psi_\varepsilon(0)$ algebraically  close.
The result for the first component reads
 \begin{widetext}
  	\begin{equation}
 	\psi_{a,\varepsilon}=\sqrt{2\lambda}e^{-i\frac{\Phi+\varphi_\varepsilon + \phi_{ext}}{4}}\frac{e^{i\frac{\varphi_\varepsilon}{2}}\left ( e^{i\frac{\phi_{ext}+\Phi}{2}}\sin{\frac{\phi_{ext}-\varphi_\varepsilon}{2}}-\lambda \sin{\frac{\Phi+\varphi_\varepsilon}{2}}\right )\chi_{l,\varepsilon} + \left (\sin{\frac{\phi_{ext}-\varphi_\varepsilon}{2}}+\lambda e^{i\frac{\phi_{ext}+\Phi}{2}} \sin{\frac{\Phi+\varphi_\varepsilon}{2}}\right )\chi_{r,\varepsilon}}{(1+\lambda^2)\cos{\varphi_\varepsilon}-\cos{\phi_{ext}}-\lambda^2\cos{\Phi} + 2i\lambda \sin{\varphi_\varepsilon}}. \label{psi_a}
 	\end{equation}
  \end{widetext}
The Dirac field in the $b$-channel is obtained using the geometrical symmetry of the setup and is given by 
  \begin{equation}
  \psi_{b,\varepsilon}(\Phi,\chi_r,\chi_l){=}\psi_{a,\varepsilon}(-\Phi,\chi_l,\chi_r). \label{psi-b}
  \end{equation}
The operator relations (\ref{psi_a}) and (\ref{psi-b}) 
are used to calculate 
$$
j=j_a-j_b=-ev\int\left(\langle\psi_{a,\varepsilon}^+\psi_{a,\varepsilon}\rangle-\langle\psi_{b,\varepsilon}^+\psi_{b,\varepsilon}\rangle\right)d\varepsilon,
$$
which results in the C$\Phi$R presented in (\ref{jj-main}).


\begin{thebibliography}{40}
	\bibitem{Alicea} J. Alicea, Rep. Prog. Phys. {\bf 75}, 076501 (2012).
	\bibitem{NayakReview}C. Nayak, S. H. Simon, A. Stern, M. Freedman, and S. Das Sarma, Rev. Mod. Phys. {\bf 80}, 1083 (2008).
	\bibitem{Kitaev}  A. Yu. Kitaev, Phys. Usp. {\bf 44} (suppl.), 131 (2001).
	\bibitem{ReadGreen} N. Read and D. Green, Phys. Rev. B {\bf 61}, 10267 (2000).
	\bibitem{MooreRead} G. Moore and N. Read, Nucl. Phys. {\bf B360}, 362 (1991).
	
	\bibitem{FuKane}L. Fu and C. L. Kane, Phys. Rev. B {\bf 79}, 161408
	(2009).
	\bibitem{OregRefaelOppen}Y. Oreg, G. Refael, and F. von Oppen, Phys. Rev.
	Lett. {\bf 105}, 177002 (2010).
	\bibitem{NilssonAkhmerov} J. Nilsson, A. R. Akhmerov, and C.W. J. Beenakker, Phys. Rev.
	Lett. {\bf 101}, 120403 (2008).
	
	\bibitem{FuKane3DTI}L. Fu and C. L. Kane, Phys. Rev. Lett. {\bf 100}, 096407 (2008).
	
	\bibitem{HasanKane}M. Z. Hasan and C. L. Kane, Rev. Mod. Phys. {\bf 82}, 3045
	(2010).
	
	\bibitem{Lutchin} R. M. Lutchyn, J. D. Sau, and S. Das Sarma, Phys. Rev.
	Lett. {\bf 105}, 077001 (2010).
	
	
	
	\bibitem{1} D. M. Badiane, M. Houzet,  and J. S. Meyer,  Phys. Rev.
	Lett. {\bf 107}, 177002 (2011).
	\bibitem{2} C. W. J. Beenakker, D. I. Pikulin, T. Hyart, H. Schomerus,
	and J. P. Dahlhaus, Phys. Rev. Lett. {\bf 110}, 017003
	(2013).
	
	
	\bibitem{FuKaneMachZehnder}L. Fu and C. L. Kane, Phys. Rev.
	Lett. {\bf 102}, 216403
	(2009).
	\bibitem{AkhmerovMachZehnder} A. R. Akhmerov, J. Nilsson, and C.W. J. Beenakker, Phys. Rev.
	Lett. {\bf 102}, 216404 (2009).
	
	\bibitem{LawFabryPerot} K. T. Law, P. A. Lee, and T. K. Ng, Phys. Rev.
	Lett. {\bf 103}, 237001 (2009).
	\bibitem{ButtikerFabryPerot}J. Li, G. Fleury, and M. B\"uttiker, Phys. Rev. B {\bf 85}, 125440 (2012).
	
	
	\bibitem{HBT} G. Str\"ubi, W. Belzig, M. S. Choi, and C. Bruder, Phys.
	Rev. Lett. {\bf 107}, 136403 (2011).

         
         \bibitem{Jiang11} L. Jiang, D. Pekker, J. Alicea, G. Refael, Y. Oreg, and F. von Oppen, Phys. Rev. Lett. {\bf 107}, 236401 (2011).

		\bibitem{LMAY}S.-P. Lee, K. Michaeli, J. Alicea, and A. Yacoby,
		Phys. Rev. Lett. {\bf 113}, 197001 (2014).
		\bibitem{Tkachov} G. Tkachov, P. Burset, B. Trauzettel, and E.M. Hankiewicz, Phys. Rev. B {\bf 92}, 045408 (2015).
	

	\bibitem{MaslovStoneGoldbartLoss} D.L. Maslov, M. Stone, P.M. Goldbart, D. Loss, Phys. Rev. B {\bf 53}, 1548 (1996).
	\bibitem{Kulik} I. O. Kulik, Zh. Eksp. Theor. Phys. {\bf 57}, 1745 (1969).
	\bibitem{S-LL-S} R. Fazio, F. W. J. Hekking, and A. A. Odintsov, Phys. Rev. B {\bf 53}, 6653 (1996).
	
	

	
	\bibitem{ChamonFreedWen}C. de C. Chamon, D. E. Freed, and X. G. Wen, Phys. Rev. B {\bf 53}, 4033 (1996).
	\bibitem{EggerGrabert}R. Egger and H. Grabert, Phys. Rev. B {\bf 58}, 10761 (1998).
	\bibitem{Alicea1} L. Fidkowski, J. Alicea, N. H. Lindner, R. M. Lutchyn, and M. P. A. Fisher, Phys. Rev. B {\bf 85}, 245121 (2012).
		\bibitem{KulikOmelyanchuk}I. O. Kulik and A. N. Omel'yanchuk, Zh. Eksp. Teor. Fiz. {\bf 68}, 2139-2148 (1975).
\end{thebibliography}
\end{document}